\documentclass[%
 reprint,
superscriptaddress,
showpacs,
 amsmath,amssymb,
 aps,
]{revtex4-1}

\usepackage{rotating}
\usepackage{stackengine}
\usepackage{graphicx}
\usepackage{color}
\usepackage{verbatim}
\usepackage{subfigure}
\usepackage{tikz}
\usepackage{tkz-graph}

\DeclareMathOperator{\diag}{diag}
\definecolor{Blue}{rgb}{0,0,1}
\definecolor{Red}{rgb}{1,0,0}
\definecolor{Black}{rgb}{0,0,0}

\newcommand{\David}[1]{{\color{Black} #1}}
\newcommand{\mw}[1]{{\color{Black} #1}}
\newcommand{\Da}[1]{{\color{Black} #1}}

\begin{document}

\preprint{APS/123-QED}

\title{Quantum state engineering in arrays of nonlinear waveguides} 

\author{David Barral} 
\email{david.barral@universite-paris-saclay.fr}
\affiliation{Centre de Nanosciences et de Nanotechnologies C2N, CNRS, Universit\'e Paris-Saclay, 10 boulevard Thomas Gobert, 91120 Palaiseau, France}
\author{Mattia Walschaers}
\affiliation{Laboratoire Kastler Brossel, Sorbonne Universit\'e, CNRS, ENS - Universit\'e PSL, Coll\`ege de France, 4 place Jussieu, F-75252 Paris, France}
\author{Kamel Bencheikh}
\affiliation{Centre de Nanosciences et de Nanotechnologies C2N, CNRS, Universit\'e Paris-Saclay, 10 boulevard Thomas Gobert, 91120 Palaiseau, France}
\author{Valentina Parigi}
\affiliation{Laboratoire Kastler Brossel, Sorbonne Universit\'e, CNRS, ENS - Universit\'e PSL, Coll\`ege de France, 4 place Jussieu, F-75252 Paris, France}
\author{Juan Ariel Levenson}
\affiliation{Centre de Nanosciences et de Nanotechnologies C2N, CNRS, Universit\'e Paris-Saclay, 10 boulevard Thomas Gobert, 91120 Palaiseau, France}
\author{Nicolas Treps}
\affiliation{Laboratoire Kastler Brossel, Sorbonne Universit\'e, CNRS, ENS - Universit\'e PSL, Coll\`ege de France, 4 place Jussieu, F-75252 Paris, France}
\author{Nadia Belabas}
\email{nadia.belabas@universite-paris-saclay.fr}
\affiliation{Centre de Nanosciences et de Nanotechnologies C2N, CNRS, Universit\'e Paris-Saclay, 10 boulevard Thomas Gobert, 91120 Palaiseau, France}

\begin{abstract}
In the current quest for efficient and experimentally feasible platforms for
implementation of multimode squeezing and entanglement in the continuous variable regime, we underpin and complement our results on the generation of versatile multimode entanglement and cluster states in nonlinear waveguide arrays presented by Barral et al., Phys. Rev. Appl. $\bf{14}$, 044025 (2020). We present detailed derivations of the equations that describe the propagation of light through this system, and then we focus on parameter regimes where these equations can be solved analytically. These analytical solutions build an intuition for the wide landscape of quantum states that are accessible through the activation of pumping, coupling and measurement schemes. Furthermore, we showcase the acquired insights by using one of the identified analytical solutions to exhibit the generation, optimization and scalability of spatial linear cluster states.
\end{abstract}

\date{November 7, 2020}
\maketitle 

\section{Introduction}\label{I}

The optical continuous variable (CV) framework is a true contender for quantum communication and quantum information processing \cite{Braunstein2005}. Demonstrations of protocols putting the field fluctuations to good use have been achieved since 1992 \cite{Ou1992, Furusawa1998, Jia2004, Coelho2009, Miwa2009, Jouguet2013} towards large-scale entangled states for quantum computing \cite{Larsen2019, Asavanant2019}. However, all these major advances were achieved with table-top experiments and for research purposes only. Integrated optics can fill the gap towards the development of real-world quantum technologies \cite{Wang2019}. Entanglement and superposition underpin the advantage of quantum protocols. Thus, an integrated synthesizer of multimode entangled states is a key component for the boost of quantum technologies. Entanglement on chip has been demonstrated for two parties \cite{Lenzini2018}, in an integrated version of the bulk optics implementations that cascade squeezers and beamsplitters \cite{Yukawa2008}. We discuss here the versatility of a monolithic device --with no bends in the active region nor specific functionalized regions--  where nonlinearity and coupling act simultaneously: the array of nonlinear waveguides (ANW) \cite{Christodoulides2003}. 

Such arrays have been used in the discrete variable domain \cite{Kruse2013, Solntsev2014} and have been proposed to achieve specific multimode states in the CV domain with $N=2$ waveguides \cite{Herec2003, Barral2017b} or scaling up the number of waveguides \cite{Rai2012, Barral2019b, Barral2020b}. Recently, the ANW has been proposed as a versatile source \mw{to engineer tailored} cluster states for measurement-based quantum computing \cite{Barral2020}. \mw{Here we provide a complementary perspective by exploring both the mathematical backbone of these results and the physical insight that can be obtained from it. We showcase how entanglement and squeezing manifests in optical modes that are produced by the ANW, and we detail possible tuning parameters, derive analytical and semi-analytical solutions, and harness them to find a good working point for the generation of linear clusters.}

Our framework is based on the choice and use of different sets of available eigenmodes of the ANW which simplify the dynamics of the system and open up new possibilities to encode quantum information. \David{The use of eigenmodes} is \mw{the key to demonstrate} tunable multipartite entanglement in the frequency domain \cite{Roslund2013, Cai2017}. We \David{establish parallels between the ANW and such frequency combs, \mw{thus} connecting spatial and frequency encoding \mw{through a joint} mathematical framework}. {We further show that full engineering of multimode squeezed states can be achieved in the spatial domain} i) by specific design of the ANW shaping the nonlinearity and coupling, and designing a suitable phase matching (as suggested before \cite{Barral2020} and further exemplified here); and externally, ii) by adjusting the pumping profile and by adapting the measurement strategies of the output fields. 

The main conceptual, practical and technological assets of the ANW in comparison with other platforms are i) the large number of degrees of freedom available that enable to reconfigure its operation, ii) a number of analytical solutions based on symmetries present in arrays of waveguides that are a guide to develop specific quantum protocols, iii) the possibility to encode quantum information in the individual mode basis or in any other basis based on linear combinations of individual modes, iv) the small footprint, from few millimeters to centimeters, and v) the simplicity of the pumping-detection optical setup that can be based on available telecom fiber-optic components. \David{Notably, quantum information encoded in the individual spatial modes can be distributed to different locations of a quantum network in a natural way, which is harder to implement in \mw{the frequency} domain \cite{Arzani2018}.}

The article is organized as follows: we first derive the governing equations of propagation and squeezing in ANW in section \ref{II}. We then detail and illustrate an inventory of the tuning parameters for squeezing and entanglement engineering in section \ref{Engineer}. Notably, we give in this section \ref{Engineer} various sets of analytical solutions that provide considerable insight on the impact of the pumping phase profile on the generation of entanglement. We build on this formalism and intuition to demonstrate multimode squeezing in section \ref{Multimode} and detail the possibilities to produce linear cluster states in ANW in section \ref{Clusters}.

\section{The array of nonlinear waveguides: possible encodings and general solutions}\label{II}

\subsection{The array of nonlinear waveguides}\label{IIA}

\begin{figure}[t]
  \centering
    \includegraphics[width=0.49\textwidth]{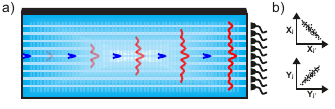}
\vspace {0cm}\,
\hspace{0cm}\caption{\label{F1}\small{a) Sketch of an array of nonlinear waveguides based on a PPLN waveguide array made up of nine waveguides working in a SPDC configuration pumping the central waveguide. Propagating pump field in blue. Evanescently-coupled SPDC signal fields in red. Quantum noise variances and correlations are measured by multimode balanced homodyne detection. b) Example of quadrature correlations between two individual SPDC signal modes $j$ and $j'$.}}
\end{figure}

The array of nonlinear waveguides consists of $N$ identical $\chi^{(2)}$ waveguides in which degenerate spontaneous downconversion (SPDC) and \Da{nearest-neighbor} evanescent coupling between the generated fields take place. The array can be made up of, for instance, periodically poled lithium niobate (PPLN) waveguides as sketched in Figure \ref{F1}a. In each waveguide, an input harmonic field at frequency $\omega_{h}$ is type-0 downconverted into a signal field at frequency $\omega_{s}$.  We consider that the phase matching condition  $\Delta\beta\equiv\beta(\omega_{h})-2\beta(\omega_{s})=0$, with $\beta(\omega_{h,s})$ the propagation constant at frequency $\omega_{h,s}$, is fulfilled all along the coupling zone and in the coupling zone only. The energy of the signal modes propagating in each waveguide is exchanged between the coupled waveguides through evanescent waves, whereas the interplay of the second harmonic waves is negligible for the considered propagation lengths due to their high confinement into the guiding region. We set our calculation in the reasonable regime of pump undepletion \cite{Barral2017b}. We consider a general array of $N$ identical waveguides and continuous-wave propagating fields. \Da{The physical processes taking place in $\chi^{(2)}$ waveguides can be described by a dynamical operator $\hat{\mathcal{M}}$ obtained quantizing the flux of momentum of the electromagnetic fields  \cite{BenAryeh1991, Toren1994}. The following Heisenberg equation is obtained for an array of $N$ evanescently coupled nonlinear waveguides in the SPDC regime \cite{Herec2003, Rai2012, Perina2000}}
\begin{equation} \label{one}
\frac{d \hat{\mathcal{A}}_{j}}{d z}=\, i C_{0} (f_{j-1} \hat{\mathcal{A}}_{j-1}+ f_{j}\hat{\mathcal{A}}_{j+1}) +2 i \eta_{j} \hat{\mathcal{A}}_{j}^{\dag}, \\
\end{equation}
where $\hat{\mathcal{A}}_{0}=0$ and $\hat{\mathcal{A}}_{N+1}=0$, $f_{0}=f_{N}=0$ and $j=1,\dots, N$ is the individual mode index. \Da{$\hat{\mathcal{A}}_{j}\equiv \hat{\mathcal{A}}_{j}(z, \omega_{s})$ are monochromatic slowly-varying amplitude annihilation operators of signal (s) photons corresponding to the $j$th waveguide --the individual mode basis-- fulfilling equal space commutation relations $[\hat{\mathcal{A}}_{j}(z,\omega), \hat{\mathcal{A}}_{j'}^{\dag}(z,\omega')]=\delta(\omega-\omega')\delta_{j,j'}$ \cite{BenAryeh1991}}. The effective nonlinear coupling constant corresponding to the $j$th waveguide is given by $\eta_{j}=g\, \alpha_{h,j}$, where $g$ is the nonlinear constant -- proportional to $\chi^{(2)}$ and to the spatial overlap of the signal and harmonic fields in each waveguide -- and $\alpha_{h,j}$ is the strong coherent undepleted pump field propagating in the $j$th waveguide. The parameters $\eta_{j}$ can be tuned by means of a suitable set of pump phases and amplitudes at each waveguide. \Da{$C_{j}=C_{0} f_{j}$ is the linear coupling constant between modes $j$ and $j+1$, where $C_{0}$ is the coupling strength and $f_{j}$ are the elements of the coupling profile $\vec{f}$}. $z$ is the coordinate along the direction of propagation. Both the coupling and nonlinear constants depend on the set signal frequency, $C_{0} \equiv C_{0}(\omega_{s})$ and $g \equiv g(\omega_{s})$, and they are taken as real without loss of generality. 

\begin{table*}[t]
\centering
\begin{tabular}{c c c}
\hline\hline
Function  &  Frequency combs \cite{Arzani2018}   &   ANWs \\[0.5ex]
\hline
Individual modes & $\hat{a}_{j}$ & $\hat{\mathcal{A}}_{j}$ \\[1ex]
Generator indiv. mod. & \qquad$\hat{\mathcal{H}}=\frac{i\hbar}{2}\sum_{j,k=1}^{N} \tilde{\mathcal{L}}_{j,j'} \hat{a}_{j}^{\dag} \hat{a}_{j'}^{\dag} + H.c.$ &  \qquad$\hat{\mathcal{M}}=\hbar  \sum_{j=1}^{N} \{C_{0} (f_{j} \hat{\mathcal{A}}_{j+1} \hat{\mathcal{A}}_{j}^{\dag}+f_{j-1} \hat{\mathcal{A}}_{j-1} \hat{\mathcal{A}}_{j}^{\dag})+ \eta_{j} \hat{\mathcal{A}}_{j}^{\dag \,2}+ H.c.\}$  \\ [1ex] 
Linear supermodes & n.a.  &$\hat{B}_{k}=\hat{\mathcal{B}}_{k}\,e^{-i\lambda_{k}z}=\sum_{j=1}^{N} M_{k,j}\,\hat{\mathcal{A}}_{j} \,e^{-i\lambda_{k}z}$ \\ [1ex] 
Generator lin. superm. & n.a.  & \qquad$\hat{\mathcal{M}}_{LS}=-\frac{i\hbar}{2}\sum_{k,k'=1}^{N} \mathcal{L}_{k,k'} (z) \hat{B}_{k}^{\dag} \hat{B}_{k'}^{\dag} + H.c.$ \\ [1ex]
Coupling matrix & $\tilde{\mathcal{L}}_{j,j'}=sinc [\phi(\omega_{j}, \omega_{j'})] \alpha(\omega_{j} + \omega_{j'})$ & $\mathcal{L}_{k,k'} (z)=2i\sum_{j=1}^{N} \vert \eta_{j}\vert M_{k,j} M_{k',j}\,e^{i\{\phi_{j}-(\lambda_{k}+\lambda_{k'})z\}}$  \\ [1ex]
Nonlinear supermodes & $\hat{c}_{m}=\sum_{j=1}^{N}V^{\dag}_{m,j} \hat{a}_{j}$ &  $ \hat{\mathcal{C}}_{m}=\sum_{k,j=1}^{N} (\Upsilon_{m,k}^{\dag}(z)\, M_{k,j}\,e^{-i\lambda_{k} z}) \hat{\mathcal{A}}_{j}$ \\[1ex]
Gen. nonl. sup. basis & $\hat{\mathcal{H}}_{NLS}=\frac{i\hbar}{2}\sum_{m=1}^{N} \Lambda_{m, m}(\hat{c}_{m}^{\dag})^{2} + H.c.$ & $\hat{\mathcal{M}}_{NLS}=\frac{i\hbar}{2}\sum_{m=1}^{N} \tilde{\Lambda}_{m,m} (z) (\hat{\mathcal{C}}_{m}^{\dag})^{2} + H.c.$ \\[1ex]
\hline
\end{tabular}
\caption{\label{Table1}\small{Comparison between SPDC in frequency combs (left) and in nonlinear waveguide arrays (right). We show for both frameworks individual modes ($\hat{a}_{j}$, $\hat{\mathcal{A}}_{j}$), dynamical generators in the individual basis ($\hat{\mathcal{H}}$, $\hat{\mathcal{M}}$), coupling matrix ($\tilde{\mathcal{L}}$, $\mathcal{L}$), respectively joint spectral and spatial distribution, nonlinear supermode basis ($\hat{c}_{m}$, $\hat{\mathcal{C}}_{m}$) and dynamical generator in the nonlinear supermode basis ($\hat{\mathcal{H}}_{NLS}$, $\hat{\mathcal{M}}_{NLS}$). Note that since the nonlinear supermodes are only defined locally, $\hat{\mathcal{M}}_{NLS}$ is a formal generator shown here for the sake of comparison. In frequency combs, the coupling between the individual modes is nonlinear and the diagonal basis of nonlinear supermodes is obtained directly from them. In contrast, in arrays of nonlinear waveguides the coupling between the individual modes is linear, which produces linear supermodes $\hat{\mathcal{B}}_{k}$. \Da{The slowly-varying linear supermodes $\hat{B}_{k}$ present a dynamical generator $\hat{\mathcal{M}}_{LS}$. The coupling between different slowly-varying linear supermodes is nonlinear and local, and the nonlinear supermodes are obtained by diagonalizing the coupling matrix of the slowly-varying linear supermodes.} n.a.: not applicable.}}
\end{table*}

Since we are interested in CV squeezing and entanglement, we will also use along the paper the field quadratures  $\hat{x}_{j}$ and $\hat{y}_{j}$, where $\hat{x}_{j}=(\hat{\mathcal{A}}_{j}+\hat{\mathcal{A}}_{j}^{\dag})$ and $\hat{y}_{j}=i (\hat{\mathcal{A}}_{j}^{\dag}-\hat{\mathcal{A}}_{j})$ are, respectively, the amplitude and phase quadratures corresponding to a signal optical mode ${\mathcal{A}}_{j}$ (Figure \ref{F1}b). The system of equations ($\ref{one}$) in terms of the individual-modes quadratures can be rewritten in compact form as 
\begin{equation}\label{Sz}
\frac{d \hat{\xi}}{d z} = {\Delta}(z)\, \hat{\xi},
\end{equation}
where ${\Delta}(z)$ is a $2N\times2N$ matrix of coefficients and $\hat{\xi}=(\hat{x}_{1},\dots, \hat{x}_{N}, \hat{y}_{1},\dots,\hat{y}_{N})^T$. 

In general, either Equation (\ref{one}) or Equation (\ref{Sz}) can be solved numerically for a specific set of parameters $(C_{j},\eta_{j},N)$, or even analytically if $N$ is small. However, it is difficult to gain physical insight from numerical or low-dimension analytical solutions due to the increasing complexity of the system with the number of waveguides. We propose below and use throughout the paper two modal approaches --complementary to the individual-mode approach--that enlighten the problem of propagation in ANWs. We thus use


i) {\it Linear (propagation) supermodes \Da{${\mathcal{B}_{k}}$}}, i.e. the eigenmodes of the corresponding linear array of waveguides assuming $\eta_{j}\propto g=0$. In the actual array of nonlinear waveguides where $\eta_{j}\propto g\neq 0$, these modes are squeezed and coupled through the nonlinearity. This basis has analytical solutions independently of the number $N$ of waveguides for specific pump-field distributions. We show these solutions in section \ref{Engineer}.

ii) {\it Nonlinear (squeezing) supermodes \Da{${\mathcal{C}_{m}}$}}, i.e. the eigenmodes of the full nonlinear system. These modes are squeezed and by construction fully decoupled but $z$-dependent. We point out that, in some cases (see section \ref{Engineerb1}), both linear and nonlinear supermodes are degenerate up to local phases. 

These two complementary approaches connect our work and our result of {spatial multimode squeezed states} exhibited in sections \ref{Multimode} and \ref{Clusters} to the spectral \cite{Roslund2013, Chen2014, Cai2017}, spatial \cite{Su2012, Armstrong2012, Armstrong2015} or temporal \cite{Yoshikawa2016, Takeda2019} modes of previous works. In the next section \ref{III}, we introduce both linear and nonlinear supermode bases, work out the corresponding propagation equations, and give the general solution to the propagation problem. Furthermore, we use the relationship between the two bases to draw mathematical parallels with SPDC frequency modes \cite{Patera2010, Arzani2018, Fabre2019}.

\subsection{Propagation equations}\label{III}

The general solutions to the propagation in ANWs have been recently introduced in \cite{Barral2020}. {Below we present a detailed calculation of those solutions in both the complex and quadratures representation of the optical fields, and compare our spatial domain solutions with those obtained in similar physical systems working with frequency modes.}

\subsubsection{Complex optical fields}\label{IIIa}

Considering coupling only between nearest-neighbor waveguides, a linear waveguide array -- i.e. Equation (\ref{one}) with $\eta_{j}=0$ -- presents \Da{linear supermodes $\hat{\mathcal{B}}_{k}$},  i.e. propagation eigenmodes \cite{Kapon1984}. In general, any linear waveguide array is represented by a Hermitian tridiagonal matrix -- Jacobi matrix -- with non-negative entries and thus by a set of non-degenerate eigenvalues and eigenvectors given in terms of orthogonal polynomials \cite{Bosse2017}. These eigenvectors, that we call the linear supermodes, form a basis and are represented by an orthogonal matrix \Da{$M\equiv M(\vec{f})$} with real elements $M_{k,j}$. The individual modes of the waveguides and the linear supermode basis are thus related by 
\begin{align}\nonumber
\hat{\mathcal{B}}_{k}=\sum_{j=1}^{N} M_{k,j}\,\hat{\mathcal{A}}_{j}.
\end{align}
\Da{These supermodes are orthonormal
\begin{equation}\label{orto}
\sum_{j=1}^{N}M_{k,j} M_{k',j}= \delta_{k,k'},
\end{equation}
with a spectrum of eigenvalues $\lambda_{k}\equiv\lambda_{k}(C_{0},\vec{f})$}. \Da{We focus on the relevant case of constant coupling along propagation, i.e. $C_{j}$ does not depend on $z$. Equation (\ref{one}) for the nonlinear waveguide array can be written as
\begin{equation}\label{three}
\frac{d {\hat{\mathcal{B}}}_{k}}{d z}=i\lambda_{k}\hat{\mathcal{B}}_{k}+2i\sum_{j=1}^{N} \sum_{k'=1}^{N} \eta_{j} M_{k,j}\,M_{k',j} \hat{\mathcal{B}}_{k'}^{\dag},
\end{equation}
in the linear supermode basis, where we have used the eigenvalue condition $C_{0}(f_{j-1} M_{k', j-1}+f_{j} M_{k',j+1})= \lambda_{k'} M_{k', j}$ and the orthogonality of the supermodes. Using slowly-varying supermode amplitudes $\hat{B}_{k}=\hat{\mathcal{B}}_{k}\,e^{-i\lambda_{k}z}$ the following propagation equation is obtained}
\begin{equation} \label{four}
\frac{d\hat{B}_{k}}{d z}=2i\sum_{j=1}^{N} \sum_{k'=1}^{N} \eta_{j}\,M_{k,j}\,M_{k',j} \hat{B}_{k'}^{\dag} e^{-i(\lambda_{k}+\lambda_{k'})z}.
\end{equation}
The momentum operator in the interaction picture which produces Equation (\ref{four}) by means of the Heisenberg equations $d\hat{B}_{k}/d z=(i/\hbar)[\hat{B}_{k},\hat{\mathcal{M}}_{LS}]$ is thus
\begin{equation} \label{five}\nonumber
\hat{\mathcal{M}}_{LS}=-\frac{i\hbar}{2}\sum_{k,k'=1}^{N} \mathcal{L}_{k,k'} (z) \hat{B}_{k}^{\dag} \,\hat{B}_{k'}^{\dag} + H.c.
\end{equation}
The coupling matrix $\mathcal{L}(z)$ is the local joint-spatial supermode distribution of the ANW and its elements are given by
\begin{equation} \label{six}
\mathcal{L}_{k,k'} (z)=2i\sum_{j=1}^{N} \vert \eta_{j}\vert M_{k,j} M_{k',j}\,e^{i\{\phi_{j}-(\lambda_{k}+\lambda_{k'})z\}}.
\end{equation}
\Da{where we have used $\eta_{j}=\vert \eta_{j} \vert e^{\i \phi_{j}}$.} $\mathcal{L}(z)$ is a complex symmetric matrix which gathers all the information about the spatial shape of the pump, i.e. amplitudes and phases in each waveguide, and the signal supermodes coupling. 

\Da{The formal solution to Equation (\ref{four}) is given by \cite{Perina2000}
\begin{equation}\nonumber
\hat{B}_{k}(z)= \exp_{\leftarrow} \left\lbrace \int_{0}^{z} \tilde{\mathcal{M}}_{LS}(z') dz'   \right\rbrace \hat{B}_{k}(0),
\end{equation}
where $\tilde{\mathcal{M}}_{LS}$ stands for a superoperator defined as $\tilde{\mathcal{M}}_{LS}(z) \cdots=(i/\hbar) [\cdots,\hat{\mathcal{M}}_{LS}(z)]$ and the symbol $\exp_{\leftarrow}$ is an exponential superoperator with increasing arguments ordered to the left. This solution evidences that in general the interaction momentum $\hat{\mathcal{M}}_{LS}$ defined in the slowly-varying basis $B$ does not commute at different positions. This makes necessary to include space-ordering corrections with significant effects in the high-gain regime --for instance, above $12$ dB squeezing in single-pass type-II PDC \cite{Christ2013}. However, these space-ordering effects can be neglected here since i) a low-gain regime --small $\vert \eta_{j}\vert$-- is crucial for individual mode entanglement, indeed the light generated in each waveguide tends to remain guided without evanescent coupling if $\vert \eta_{j} \vert > C_{0}$ \cite{Fiurasek2000}, ii) our model is limited to small values of $C_{0}$, as large values should include next-to-nearest neighbor evanescent coupling where linear supermodes are not available \cite{Kapon1984}, and iii) we study SPDC generated from vacuum where space-ordering corrections start at the third order --roughly as $\mathcal{O}(\vert \eta_{j} \vert^{3})$ \cite{Quesada2014}. We emphasize that all the analytical solutions presented in section \ref{Engineerb} are nevertheless exact for any gain regime as they are also obtained in the basis $\mathcal{B}$ from Equation (\ref{three}). Thus, in the entanglement regime, Equation (\ref{four}) can be simply written as
\begin{equation}\label{Hei}
\frac{d\hat{B}_{k}}{d z}=\sum_{k'=1}^{N} \mathcal{L}_{k,k'} (z) \,\hat{B}_{k'}^{\dag},
\end{equation}
with the following formal solution
\begin{equation} \label{Bsol}
\begin{small}
\begin{pmatrix}
\vec{B}(z) \\
\vec{B}^{\dag}(z)
\end{pmatrix}= \exp \left\lbrace 
\begin{pmatrix}
0 & \int_{0}^{z} \mathcal{L}(z') dz' \\
\int_{0}^{z} \mathcal{L}^{*}(z') dz' & 0
\end{pmatrix}
  \right\rbrace
  \begin{pmatrix}
\vec{B}(0) \\
\vec{B}^{\dag}(0)
\end{pmatrix},
\end{small}
\end{equation}
where $\vec{B}=(\hat{B}_{1}, \dots, \hat{B}_{N})^{T}$.} The solution (\ref{Bsol}) displays the effect of the nonlinearity on the linear supermodes: $z$-dependent amplification and coupling. A simpler solution can be obtained diagonalizing the matrix argument of the exponential in Equation (\ref{Bsol}) through the nonlinear supermodes. In general, the linear supermode basis does not diagonalize the propagation in the ANWs and the solution of Equation (\ref{Bsol}) is configuration-dependent. \Da{However, Equation (\ref{Hei}) --and (\ref{three})-- presents analytical solutions independently of the dimension $N$ for specific pump-field distributions as we show in section \ref{Engineer}. }

A feature of the ANWs is that the evanescent coupling produces a phase mismatch between the pump and the generated signal waves which results in a $z$-dependent interaction, in such a way that the eigenmodes of the full nonlinear system --the nonlinear supermodes-- are local. This coupling-based phase mismatch affects the amount of squeezing and entanglement generated in the ANWs. The local nonlinear supermode basis displays independently squeezed modes and helps to quantify the amount of nonclassicality generated in the array at different propagation distances. The local nonlinear supermodes basis $\mathcal{C}$ is such that
\begin{equation}\label{eight}
 \hat{\mathcal{C}}_{m}=\sum_{k=1}^{N} \Upsilon_{m,k}^{\dag}(z)\,\hat{B}_{k},
\end{equation}
where \Da{$[\hat{\mathcal{C}}_{m}(z,\omega),\hat{\mathcal{C}}_{m'}^{\dag}(z,\omega')]=\delta(\omega-\omega')\delta_{m,m'}$}, and $\Upsilon(z)$ is an unitary matrix which diagonalizes the complex symmetric matrix $\int_{0}^{z} \mathcal{L}(z') dz'$ by a congruence transformation -- the Autonne-Takagi transformation \cite{Cariolaro2016}--, such that
\begin{equation}\label{seven}
\Upsilon(z) \,[\int_{0}^{z} \mathcal{L}(z') dz' ] \,\Upsilon^{T}(z)= \Lambda(z),
\end{equation}
with $\Lambda(z)$ a local diagonal matrix with non-negative real entries.  Applying Equations (\ref{eight}) and (\ref{seven}) on Equation (\ref{Bsol}) we obtain a simple solution in the diagonal local basis 
\begin{align} \label{nineC}
\hat{\mathcal{C}}_{m}(z)=\cosh[{r}_{m}(z)] \,\hat{\mathcal{C}}_{m}(0)+\sinh[{r}_{m}(z)]\,\hat{\mathcal{C}}_{m}^{\dag}(0).
\end{align}
Each local nonlinear supermode is a single-mode squeezed state. The $r_{m}(z)=\Lambda_{m,m}(z)$ are the downconversion gains at a propagation distance $z$ and quantify the available nonlinearity and thus squeezing. The relation between the nonlinear supermodes and the individual modes is
\begin{equation}\label{NSup}\nonumber
 \hat{\mathcal{C}}_{m}=\sum_{k=1}^{N} \sum_{j=1}^{N} (\Upsilon_{m,k}^{\dag}(z)\, M_{k,j}\,e^{-i\lambda_{k} z}) \hat{\mathcal{A}}_{j} .
 \end{equation}
This expression encapsulates the mechanisms at play in the ANWs: the evanescent coupling generates the linear supermodes ($M_{k,j}$) which get a phase due to propagation ($\lambda_{k} z$) and the nonlinearity couples them locally ($\Upsilon_{m,k}^{\dag}(z)$). In terms of the individual modes, the solution to the nonlinear system is
\begin{align}\nonumber
\hat{\mathcal{A}}_{j}(z)=&\sum_{k, m, j'=1}^{N} (M_{k,j} \Upsilon_{m,k}(z) M_{m,j'}\, e^{i \lambda_{k} z}) \\ \label{IndGenSol}
&\{\cosh[r_{m}(z)]\,\hat{\mathcal{A}}_{j'}(0)+\sinh[r_{m}(z)] \,\hat{\mathcal{A}}_{j'}^{\dag}(0)\}.
\end{align}
Equations (\ref{Bsol}), (\ref{nineC}) and (\ref{IndGenSol}) are the general solutions \Da{in the low-gain regime} to the propagation problem in ANW in the linear supermodes, nonlinear supermodes and individual mode bases, respectively. These three solutions represent a resource for encoding quantum information. Particularly, Equation (\ref{IndGenSol}) is a useful tool in the DV framework to explore further, for instance, driven quantum walks \cite{Hamilton2014}.

Remarkably, the kind of equations that we find here for spatial modes are formally similar to those that appear in the context of SPDC in frequency combs \cite{Patera2010, Arzani2018, Fabre2019}. We draw parallels between the spectral approach that leads to multimode entanglement and our spatial approach in ANW in Table \ref{Table1}. In frequency combs, the individual modes are a discrete set of $N$ frequency modes $\hat{a}_{j}$ that are nonlinearly coupled in a bulk crystal with a quadratic nonlinearity. The diagonalization of the corresponding coupling matrix $\tilde{\mathcal{L}}$ produces a set of nonlinear supermodes $\hat{c}_{m}$, whose eigenvalues $\Lambda_{m, m}$ are proportional to SPDC gains. Table \ref{Table1} (left) shows the main elements involved in frequency-comb SPDC and the related Hamiltonian in the individual $\hat{\mathcal{H}}$ and nonlinear supermode $\hat{\mathcal{H}}_{NLS}$ bases. In ANWs the evanescent coupling between the individual modes $\hat{\mathcal{A}}_{j}$ generates the linear supermodes $\hat{\mathcal{B}}_{k}$ and the nonlinear coupling mediated by the pump fields mixes them. Table \ref{Table1} (right) shows the main elements involved in spatial ANWs and the related momenta in the individual $\hat{\mathcal{M}}$, linear supermode basis $\hat{\mathcal{M}}_{LS}$, and nonlinear supermode $\hat{\mathcal{M}}_{NLS}$ basis. Note that $\hat{\mathcal{M}}_{NLS}$ does not represent a real dynamical generator. It is indeed a formal squeezing momentum with singular values $ \tilde{\Lambda}_{m,m}(z)=d_{z} \Lambda_{m,m}(z)$ defined only at a set $z$ that we define for the sake of comparison. The coupling matrix $\mathcal{L}(z)$ is defined here in the linear supermode basis and the diagonalization of $\int_{0}^{z} \mathcal{L}(z') dz' $ produces a set of nonlinear supermodes $\hat{\mathcal{C}}_{m}$. The consequence of this diagonalization in two steps to get to nonlinear coupling (see Table \ref{Table1}) is that the nonlinear supermodes are $z$-dependent -- i.e. local. At each propagation plane $z$ a different set of nonlinear supermodes diagonalizes Equation (\ref{Bsol}) with SPDC gains $r_{m}(z)$. This feature is the main conceptual difference between frequency combs and spatial ANWs. It makes the ANW system very complex but, equally, highly versatile for the generation of multimode quantum states.

\subsubsection{Quadratures of the optical fields}\label{IIIb}

In terms of individual modes quadratures  $\hat{x}_{j}$ and $\hat{y}_{j}$, the full evolution of the system is obtained by solving Equation (\ref{Sz}). The formal solution of this equation is given by 
\begin{equation}\nonumber
\hat{\xi}(z)={S}(z)\, \hat{\xi}(0),
\end{equation}
with ${S}(z)=\exp\{\int_{0}^{z}{\Delta}(z')\, d z' \}$. The propagator ${S}(z)$ is a symplectic matrix which contains all the information about the propagation of the quantum state of the system. We can apply on it a Bloch-Messiah decomposition as follows \cite{Braunstein2005a}
\begin{equation}\nonumber
{S}(z)=R_{1}(z) K(z) R_{2}(z),
\end{equation}
where $R_{1}(z)$ and $R_{2}(z)$ are both orthogonal and symplectic matrices and $K(z)=\diag\{e^{r_{1}(z)}, e^{r_{2}(z)},\dots, e^{r_{N}(z)}, e^{-r_{1}(z)}, e^{-r_{2}(z)}, \dots, e^{-r_{N}(z)}\}$ is a phase-squeezed diagonal matrix. \Da{The nonlinear supermodes obtained in this way are the same as those of Equation (\ref{eight}) by the Autonne-Takagi factorization, and the squeezing factors $r_{m}(z)$ are the same as the downconversion gains obtained in the previous section \cite{Cariolaro2016}.} The spatial profiles of the nonlinear supermodes are obtained from the complex representation of $R_{1}$.

The quantum states generated in ANWs are Gaussian. The most interesting observables in Gaussian CV are the second-order moments of the quadrature operators, properly arranged in the covariance matrix ${V}$ \cite{Adesso2014}. The elements of this matrix can be efficiently measured by means of homodyne detection. For a quantum state initially in vacuum, the covariance matrix at any plane $z$ is given by ${V}(z)={S}(z)\,{S}^{T}(z)$, with 1 the value of the shot noise related to each quadrature in our notation. Evolution of variances $V(\xi_{i}, \xi_{i})$ and quantum correlations $V(\xi_{i}, \xi_{j})$ can be obtained at any length from the elements of this matrix. The covariance matrix can also be computed from the Bloch-Messiah decomposition as 
\begin{equation}\label{BMD}
V(z)={R}_{1}(z) K^{2}(z){R}^{T}_{1}(z).
\end{equation} 
Thus, $K^{2}(z)$ is the covariance matrix in the nonlinear supermode basis and $R_{1}(z)$ the symplectic transformation matrix between the individual and nonlinear supermode basis [equivalent to Equation (\ref{NSup}) for complex fields]. The $m$th nonlinear supermode is squeezed and thus nonclassical if $K^{2}_{N+m}(z)= e^{-2 r_{m}(z)}<1$, and the smallest value of $K^{2}_{N+m}(z)$ is called the generalized squeezed variance and it is a measure of the nonclassicality of the quantum state \cite{Simon1994}. 

Note that the complex and real approaches are equivalent \cite{Cariolaro2016}. The first method is applied to the complex joint-spatial supermode distribution and is numerically easier to compute. It gives the relative downconversion gains and therefore the amount of squeezing available in the ANW. The second method is applied to the propagator in the symplectic form and it enables to work out directly the noise properties of the quantum state. We detail how these noise properties can be engineered in the following sections.

\section{Engineering toolbox for production and detection of multimode squeezing}\label{Engineer}

To operate the ANW, several knobs are accessible experimentally. A reconfigurable multimode shaper at pump frequency inputs the desired profile ($\vec{\eta}, \vec{\phi}$) in the array through a V-groove fiber array. Bent waveguides conduct the pump modes to the periodically poled ANWs  where signal modes are generated and evanescently coupled. The coupling profile, wavevector phasematching poling period and coupling phasematching poling period ($\vec{f}, \Lambda_{\Delta\beta}, \Lambda_{C}$) can be suitably engineered for a specific operation mode. The output light is collected by  V-groove fiber arrays and directed to a multimode balanced homodyne detector (BHD) where modes are measured using adapted local-oscillator (LO) phase and electronic gain profiles ($\vec{\theta}, \vec{G}$).

The class of ANWs which we introduced in section \ref{IIA} thus presents a number of parameters that can be engineered for a desired operation. The evanescent coupling profile $\vec{f}=(f_{1},\dots ,f_{N})$, the length of the sample $L$, the number of waveguides $N$ --and notably its parity-- and the poling periods \cite{Barral2019c} are built-in and cannot be tuned once the sample is fabricated. In contrast, the power and phase pump profile, given respectively by $\vec{\eta}=(\vert\eta_{1}\vert, \dots, \vert\eta_{N}\vert)$ and $\vec{\phi}=(\arg{(\eta_{1})}, \dots, \arg{(\eta_{N})})$, the coupling strength $C_{0}$ and the basis of detection can be set for a required operation or encoding of information. $C_{0}\equiv C_{0}(\omega_{s})$ can indeed be adjusted by tuning the phase matching $\Delta \beta \equiv \Delta \beta (T)$ with the temperature $T$ of the sample and adjusting the frequency $\omega_{p}$ of the pump laser accordingly to recover the degeneracy point \cite{Vainio2011}. We introduce below a number of engineering strategies related to the coupling, pumping, phase matching and detection parameters that can be used to produce and detect a desired multimode squeezed state:

i) In section \ref{Engineera} we review and extend the analytical expressions for the linear supermodes and the propagation constant for three specific coupling $\vec{f}$ profiles. These coupling profiles are put to good use in section \ref{Multimode} to exemplify different squeezing behaviors. 

ii) In section \ref{Engineerb} we establish analytical joint spatial supermode \David{distributions} for specific pumping ($\vec{\eta}, \vec{\phi}$) configurations, we deduce in three \David{limiting} cases analytical expressions for the covariance matrices that are valid for any ANWs --any $\vec{f}$--, for any number of waveguides $N$ and any propagation distance $z$. We display covariance matrices in these \David{limiting} cases and in an intermediate \David{scenario}. We further use and comment these propagation results on the modes when discussing the generation of squeezing and entangled states in section \ref{Multimode}.

iii) In section \ref{Engineerc} we describe engineering of the propagation solutions using dedicated phase matching in ANW ($\Lambda_{C}$) to favor specific supermodes towards entanglement generation. 

iv) In section \ref{Engineerd} we recall LO shaping in multimode BHD used to detect squeezing and entanglement in a given basis.

\subsection{Coupling profile engineering}\label{Engineera}

As introduced section \ref{III}, every set of nearest-neighbor coupled waveguides has a family of propagation supermodes given by a matrix $M$. The slowly varying amplitude corresponding to the $k$th supermode propagates along the array with a propagation constant $\lambda_{k}$. Each family of linear supermodes depends on the coupling profile $\vec{f}$. The engineering of this profile enables a specific operation or logic gate \cite{Moison2009}. A number of demonstrations with optical lattices has been exhibited over the last years \cite{Keil2011, Chapman2016, Weimann2016}. Very recently, the production of topologically protected quantum states in a Su-Schrieffer-Heeger lattice has been demonstrated \cite{Blanco2018}. 

A summary of properties of the supermodes can be found in ref. \cite{Efremidis2005}. Particularly, every family of supermodes corresponding to an array of identical waveguides fulfill the following relations
\begin{align}\label{Rel1}
\lambda_{k}&=-\lambda_{N+1-k},\\ \label{Rel1a}
M_{N+1-k,j}&=(-1)^{j+1} M_{k,j}.
\end{align}
We label the supermodes connected two by two by Equation (\ref{Rel1}) as side supermodes ($k, N+1-k$). In arrays with an odd number of waveguides there is also a central supermode $k=(N+1)/2\equiv l$ with propagation constant $\lambda_{l}=0$. We thus refer to it as the zero supermode.

Applying the above relations in the orthonormalization condition Equation (\ref{orto}), we find the following modified orthonormality conditions
\begin{align}
\label{Rel1b}
\sum_{j=1}^{N} (-1)^{j+1} M_{k,j} &M_{k',j}=\delta_{k,N+1-k'}, \\ \label{Rel1c}
\sum_{2\leq 2j \leq N} M_{k,2j} M_{k',2j}=&\frac{1}{2}(\delta_{k, k'} - \delta_{k,N+1-k'}),\\ \label{Rel1d}
\sum_{1\leq 2j-1\leq N} M_{k,2j-1} M_{k',2j-1}=&\frac{1}{2}(\delta_{k, k'} + \delta_{k,N+1-k'}).
\end{align}
These relations are general and, notably, they are instrumental to configure the pump to obtain simple analytical solutions through Equation (\ref{six}). We derive and give such solutions in section \ref{Engineerb}. 

We exhibit below three paradigmatic examples of coupling profile engineering: the homogeneous profile array, the parabolic profile array and the square root profile array. We display the supermodes that each array produces and their respective propagation constants.

\subsubsection{Homogeneous profile array}\label{Engineera1}
The homogeneous linear array exhibits a constant coupling between waveguides $f_{j}=1$. It is thus a symmetric lattice. The supermodes are orthonormal Chebyshev polynomials that can be written in terms of simple trigonometric functions as \cite{Meng2004}
\begin{equation}\nonumber
M_{k,j}=M_{j,k}\equiv\frac{\sin(\frac{jk\pi}{N+1})}{\sqrt{\sum_{j'=1}^{N}\sin^{2}(\frac{j' k\pi}{N+1})}}.
\end{equation}
The Chebyshev supermodes for $N=5$ waveguides are sketched in Figure \ref{F1a}a. The spectrum of its eigenvalues is given by 
\begin{equation} \nonumber
\lambda_{k}=2 C_{0} \cos(\frac{k \pi}{N+1}), 
\end{equation}
which are the propagation constants related to each supermode. 

\subsubsection{Parabolic profile array}\label{Engineera2}
The parabolic linear array exhibits a coupling between waveguides given by the profile $f_{j}=\sqrt{j(N-j)}/2$. It is a symmetric lattice. The supermodes are orthonormal Krawtchouk polynomials that can be written in terms of Jacobi polynomials as \cite{Perez2013, Bosse2017} 
\begin{align} \nonumber
M_{k,j}&=2^{(j-\frac{N+1}{2})} \sqrt{\frac{(j-1)! (N-j)!}{(k-1)! (N-k)!}} P_{j-1}^{N-k+1-j,k-j}(0)\\  \nonumber
&=M_{j,k}.
\end{align}
The Krawtchouk supermodes for $N=5$ waveguides are sketched in Figure \ref{F1a}b. Note that for a small number of waveguides the Krawtchouk and Chebyshev supermodes are very similar. The eigenvalues are in this case equally spaced as given by
\begin{equation}\nonumber
\lambda_{k} = \frac{N-2k+1}{2}\,C_{0}.
\end{equation}
The continuous limit ($N \rightarrow \infty$) of these discrete eigenfunctions are the Hermite-Gaussian functions \cite{Weimann2016}. Remarkably, a parametric generalization of this set of supermodes, so-called para-Krawtchouk supermodes, allows fractional revivals and thus generalizes beam splitters --or directional couplers-- to N dimensions \cite{Bosse2017}.

\subsubsection{Square-root profile array}\label{Engineera3}
\begin{figure*}[t]
{\raisebox{0.2cm}{\begin{turn}{90} Homogeneous \end{turn} \hspace{0.2cm}}}
{\subfigure{\includegraphics[width=0.165\textwidth]{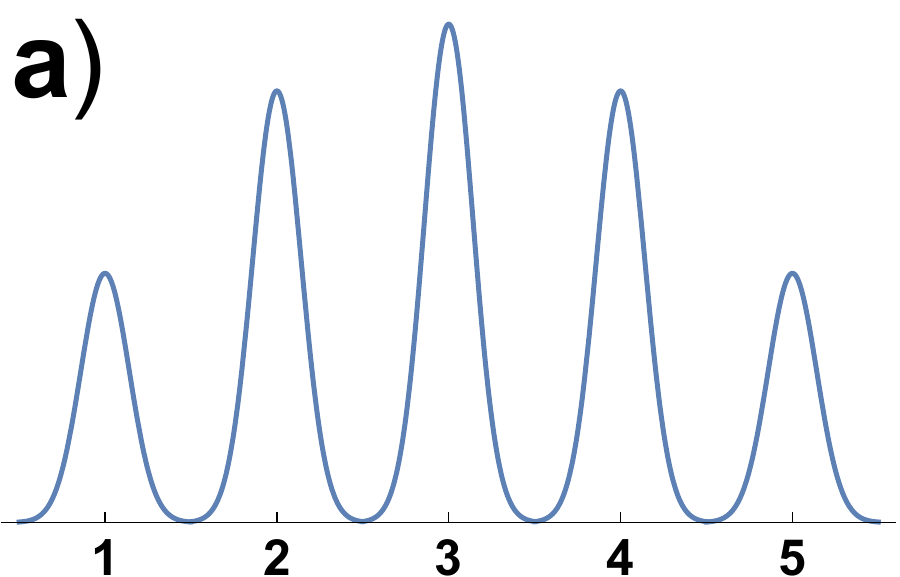}}} \hspace{0.4cm}
{\subfigure{\includegraphics[width=0.165\textwidth]{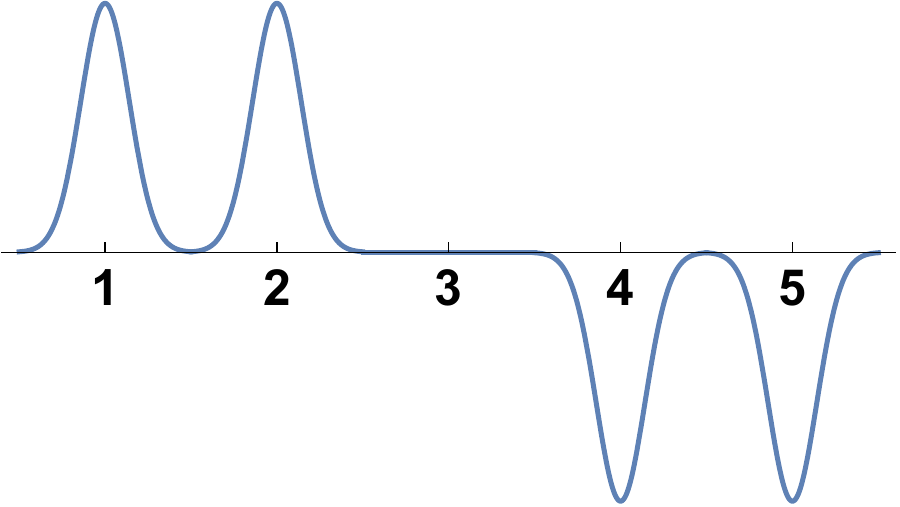}}} \hspace{0.4cm}
{\subfigure{\includegraphics[width=0.165\textwidth]{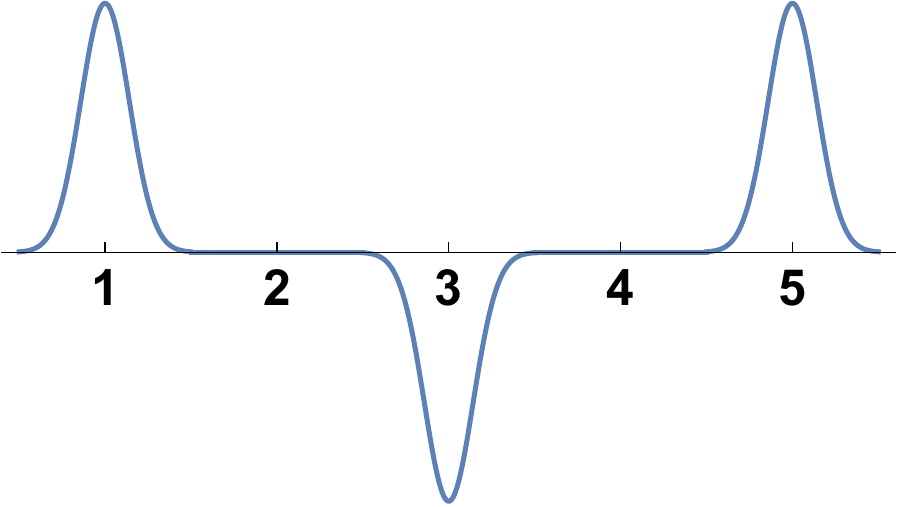}}} \hspace{0.4cm}
{\subfigure{\includegraphics[width=0.165\textwidth]{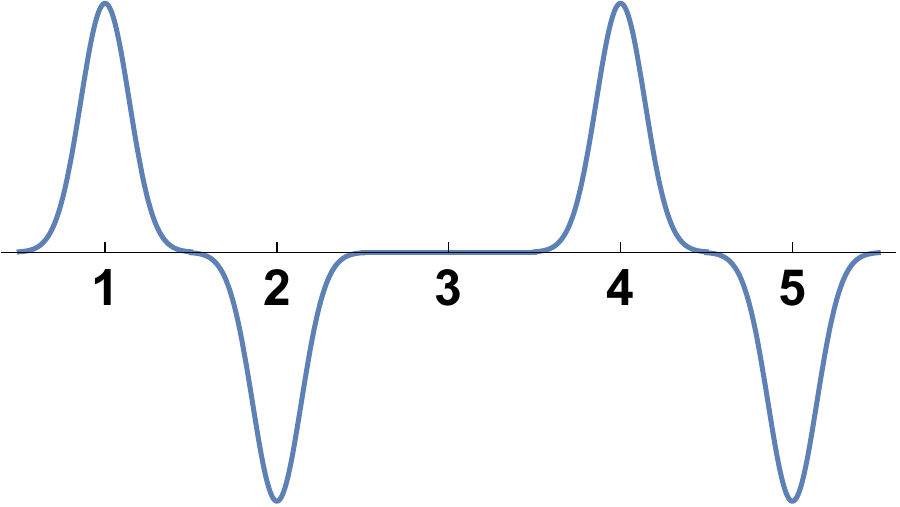}}}\hspace{0.4cm}
{\subfigure{\includegraphics[width=0.165\textwidth]{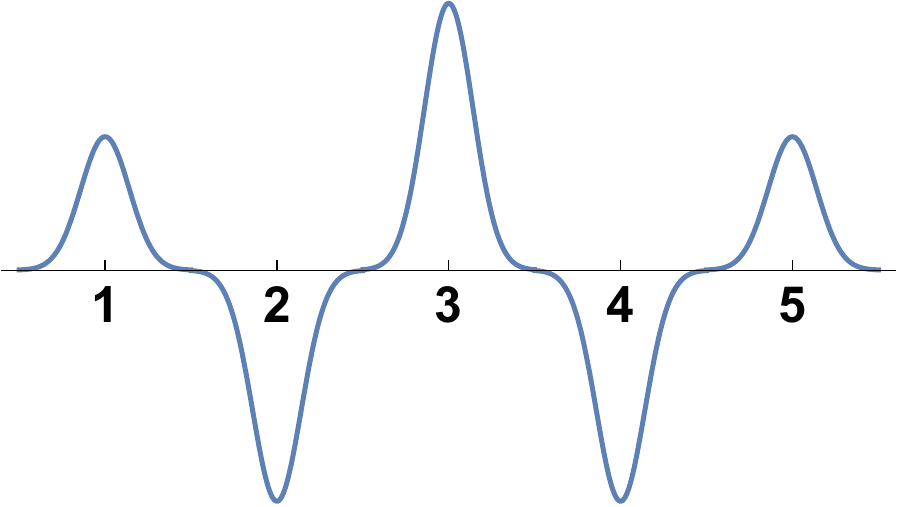}}}\\
\vspace{0.3cm}
\hspace{-0.38cm}
{\raisebox{0.2cm}{\begin{turn}{90} Parabolic \end{turn} \hspace{0.2cm}}}
\subfigure{\includegraphics[width=0.165\textwidth]{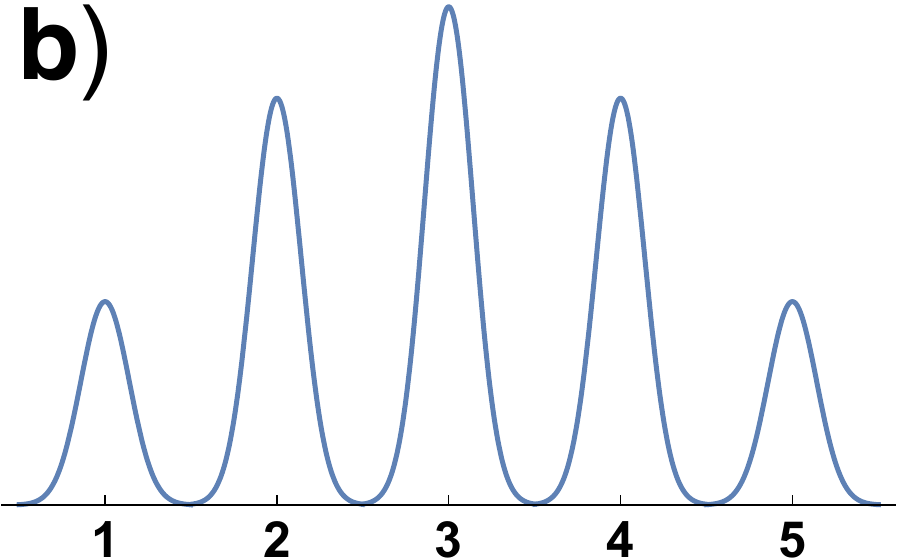}}\hspace{0.4cm}
\subfigure{\includegraphics[width=0.165\textwidth]{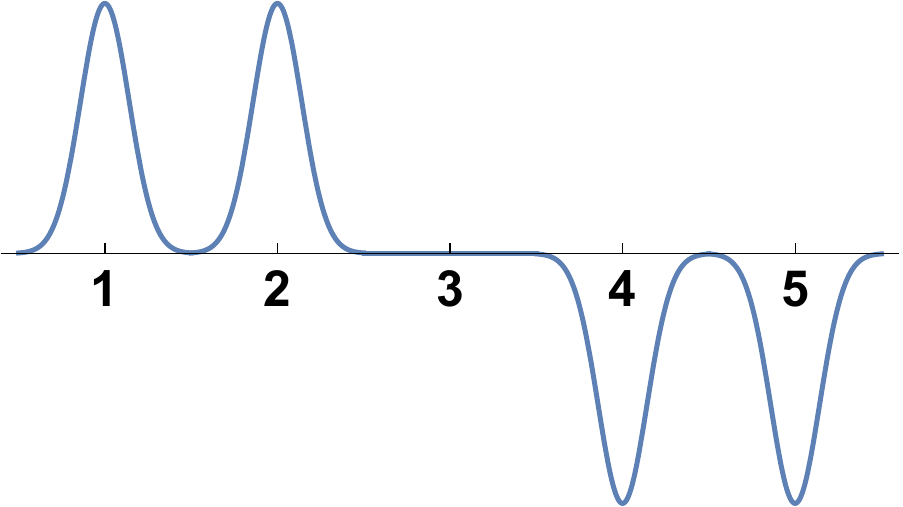}}\hspace{0.4cm}
\subfigure{\includegraphics[width=0.165\textwidth]{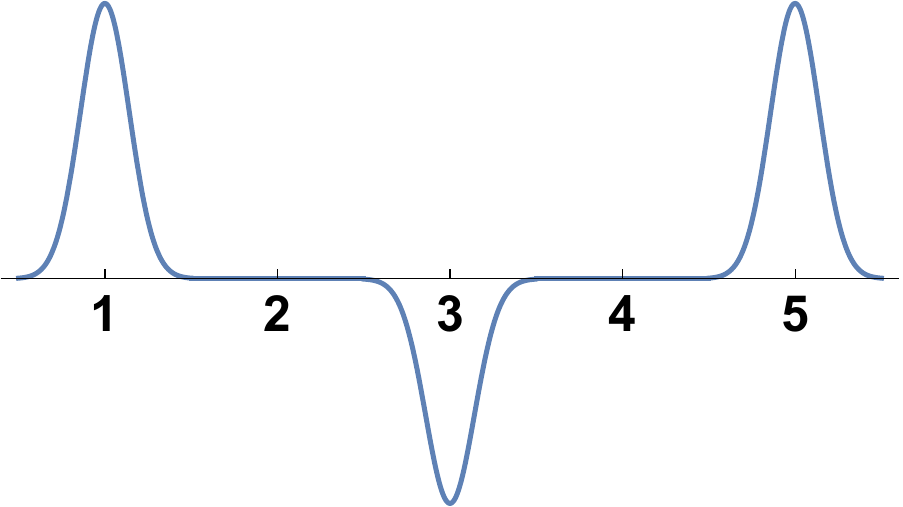}}\hspace{0.4cm}
\subfigure{\includegraphics[width=0.165\textwidth]{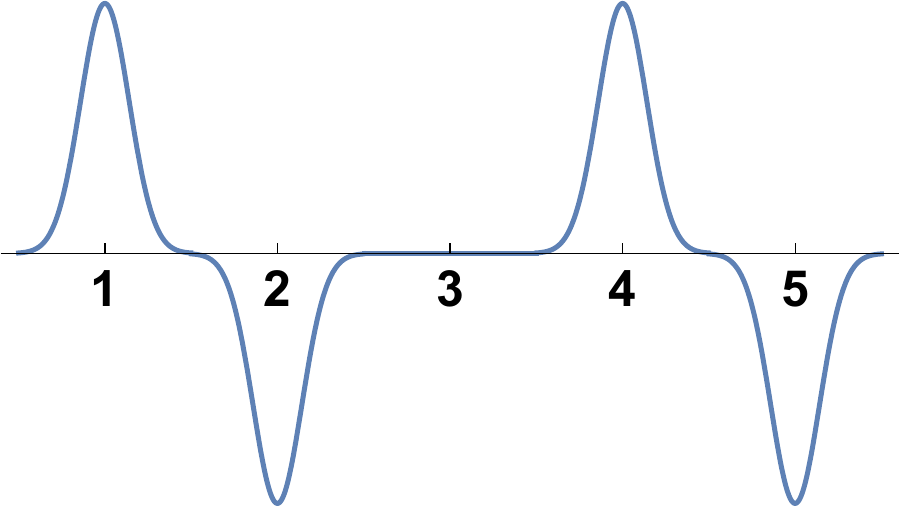}}\hspace{0.4cm}
\subfigure{\includegraphics[width=0.165\textwidth]{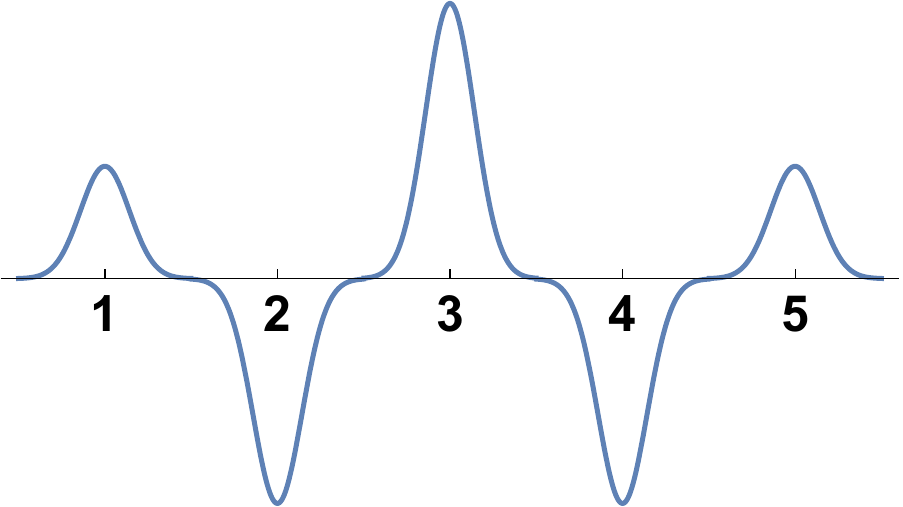}}\\
\vspace{0.3cm}
\hspace{0.cm}
{\raisebox{0.2cm}{\begin{turn}{90} Square root \end{turn} \hspace{0.2cm}}}
\stackunder[8pt]{\subfigure{\includegraphics[width=0.165\textwidth]{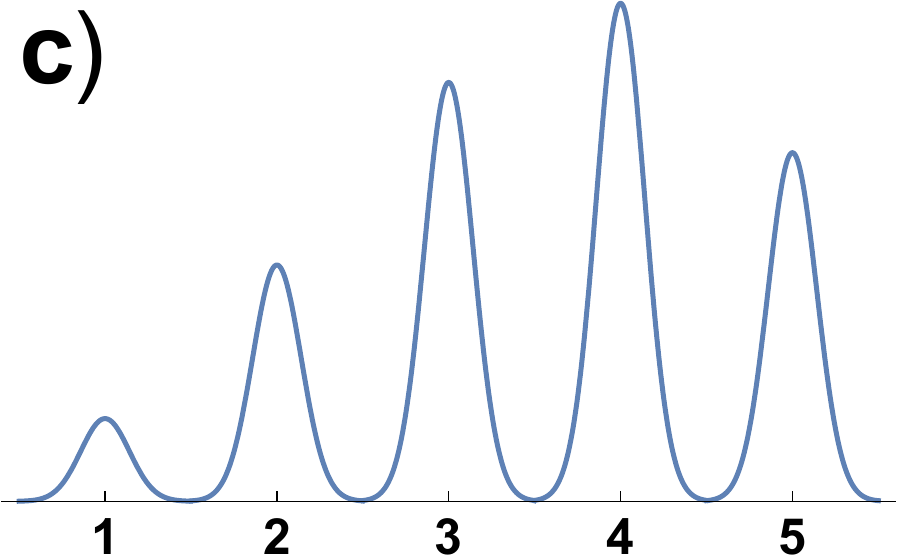}}}{${\bf k=1}$}\hspace{0.4cm}
\stackunder[8pt]{\subfigure{\includegraphics[width=0.165\textwidth]{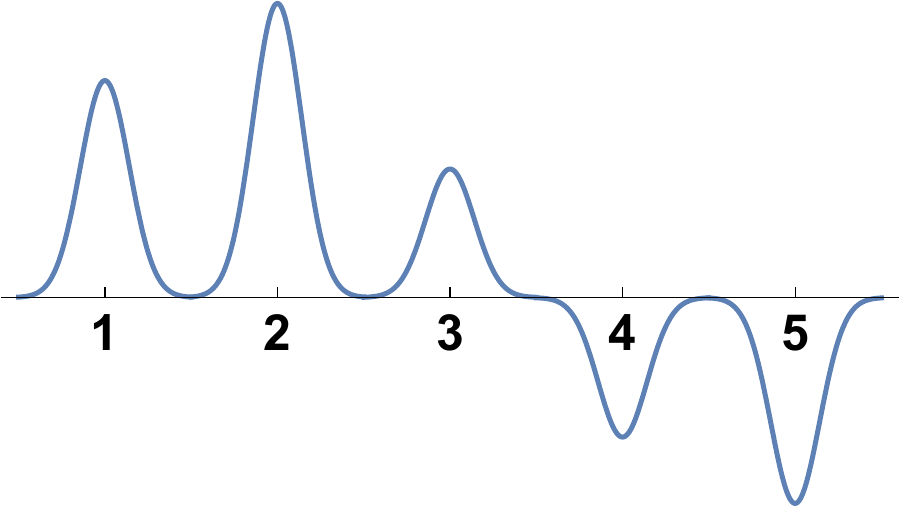}}}{${\bf k=2}$}\hspace{0.4cm}
\stackunder[8pt]{\subfigure{\includegraphics[width=0.165\textwidth]{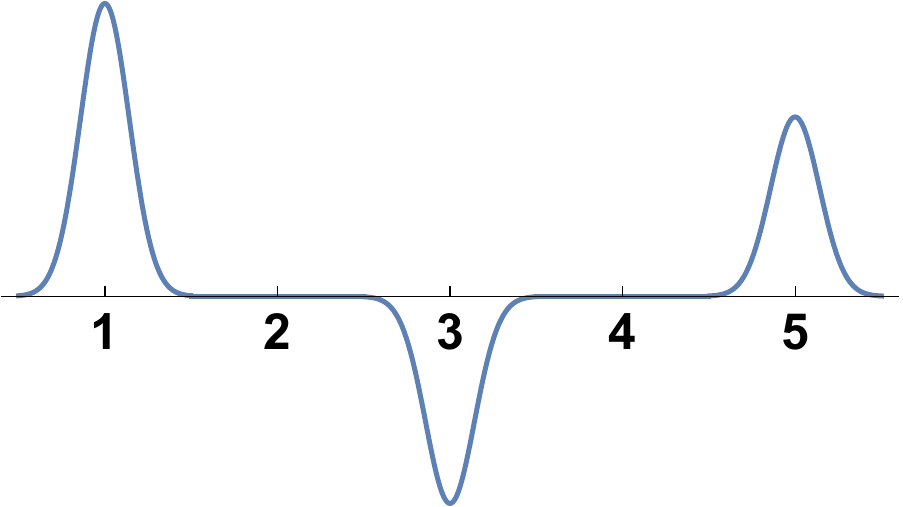}}}{${\bf k=3}$}\hspace{0.4cm}
\stackunder[8pt]{\subfigure{\includegraphics[width=0.165\textwidth]{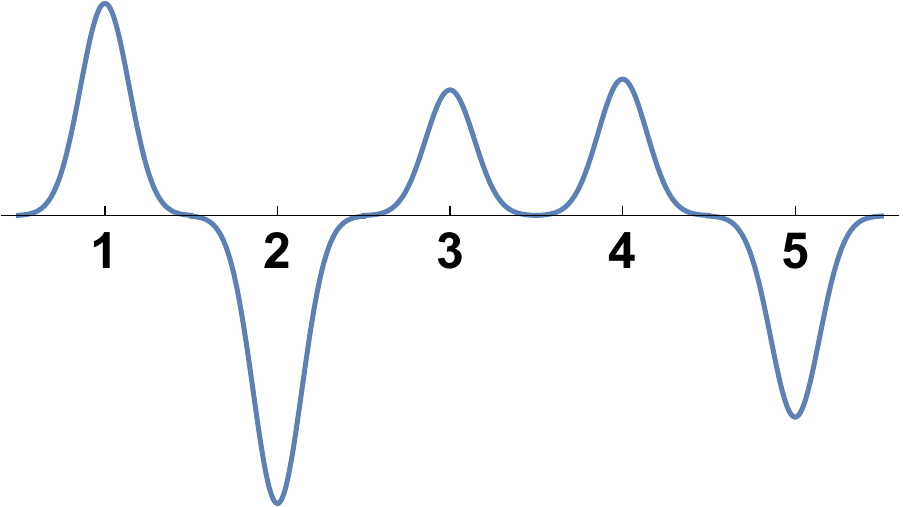}}}{${\bf k=4}$}\hspace{0.4cm}
\stackunder[8pt]{\subfigure{\includegraphics[width=0.165\textwidth]{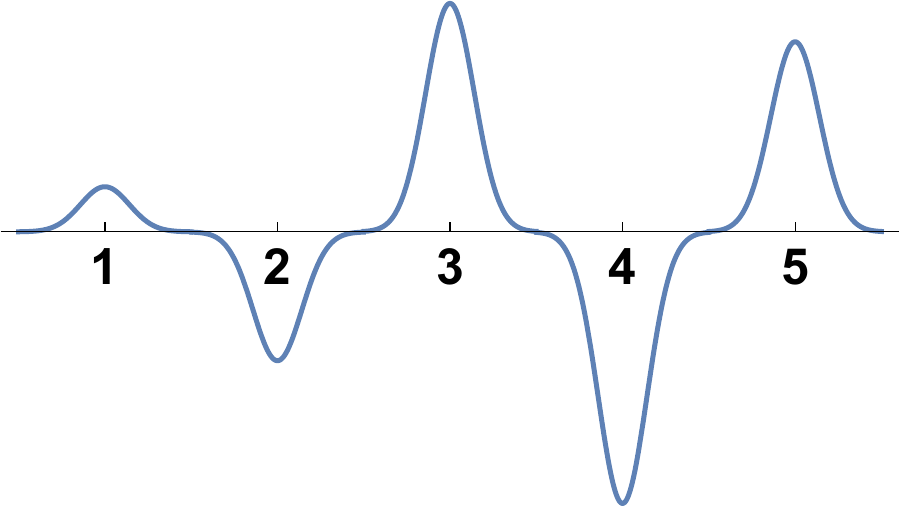}}}{${\bf k=5}$} 
\vspace {0cm}\,
\hspace{0cm}\caption{\label{F1a}\small{Sketch of the Chebyshev, Krawtchouk and Glauber-Fock supermodes related to arrays of linear waveguides with a) homogeneous, b) parabolic and c) square-root coupling profiles and $N=5$ waveguides. The horizontal axis stands for the individual modes. The propagation constants corresponding to each supermode are a) $\lambda=\{\sqrt{3}\, C_{0}, C_{0}, 0, -C_{0}, -\sqrt{3}\,C_{0}\}$, b) $\lambda=\{ 2\,C_{0}, C_{0}, 0, -C_{0}, -2C_{0}\}$ and c) $\lambda=$  $\{ \sqrt{5+\sqrt{10}}\,C_{0},$ $\sqrt{5-\sqrt{10}}\,C_{0},$ $0,$ $-\sqrt{5-\sqrt{10}}\,C_{0},$ $-\sqrt{5+\sqrt{10}}\,C_{0}\}$. $k\equiv l=3$ are the zero supermodes related to each array.}}
\end{figure*}

The square-root or Glauber-Fock linear array exhibits a coupling between waveguides given by the profile $f_{j}=\sqrt{j}$. It is an asymmetric lattice that can be symmetrized. The eigenvalues $\lambda_k$ are obtained as the roots of the $N$th Hermite polynomial given by \cite{Leija2010, Rodriguez2011}
\begin{equation} \nonumber
H_{N}[\lambda_{k'}/(\sqrt{2} C_{0})] = 0,
\end{equation}
with $k'\equiv k-1=0, \dots N-1$. The Glauber-Fock supermodes can be written in terms of normalized Hermite polynomials evaluated at these roots 
\begin{align} \nonumber
M_{k,j}=\frac{H_{j-1}(\frac{\lambda_{k'}}{\sqrt{2} C_{0}  } )}{\sqrt{2^{j-1} \,(j-1)!\, \tilde{N}_{k'}}},
\end{align}
where $\tilde{N}_{k'}=\sum_{j'=0}^{N-1} H_{j'}[\lambda_{k'}/(\sqrt{2} C_{0})]^{2}/(2^{j'} (j')!)$. The Glauber-Fock supermodes for $N=5$ waveguides are sketched in Figure \ref{F1a}c. Emulation of a driven quantum harmonic oscillator has been demonstrated in this lattice \cite{Keil2012}.

\subsection{Pump profile engineering}\label{Engineerb}

Suitable manipulation of individual power and phase pump fields by means of off-the-shelf elements as fiber attenuators and phase shifters, followed by input into the ANWs through V-groove arrays, enables an on-demand pump distribution engineering. 

The pump profile couples the propagation supermodes generating the joint-spatial supermode distribution Equation (\ref{six}). In general, this generates complicated connections between the linear supermodes. However, the orthogonality and symmetry properties of the linear supermodes [Equations (\ref{orto}) and (\ref{Rel1}) to (\ref{Rel1d})] lead to simple analytical solutions in some cases. An outstanding simplification of the system is obtained when pumping all the waveguides with the same power $\vert \eta_{j}\vert=$ constant. From now on, we refer to this as a flat pump profile. Another simplified solution is obtained when pumping only the even or odd waveguides, or when pumping only the central waveguide in an odd ANWs. Below we give the joint-spatial supermode distributions obtained with these input configurations \Da{and the exact analytical solutions to the Heisenberg Equations (\ref{Hei}) --or equally (\ref{three})-- in the simplest cases.}

\subsubsection{Flat pump profile: uniform phase}\label{Engineerb1}

When all waveguides are equally pumped such that $\vert\eta_{j}\vert=\vert\eta\vert=\tilde{\eta}$ and $\phi_{j}=\phi$, the local joint-spatial supermode distribution Equation (\ref{six}) is notably simplified to 
\begin{equation} \label{L1}\nonumber
\mathcal{L}_{k,k'} (z)=2i\, \delta_{k,k'} e^{i \{\phi-(\lambda_{k}+\lambda_{k'})z\}},
\end{equation}
where we have used the orthonormality of the linear supermodes Equation (\ref{orto}). \Da{This pump configuration diagonalizes the momentum in the slowly-varying linear supermode basis and the following Heisenberg equations are obtained
\begin{equation} \label{sixB}\nonumber
\frac{d\hat{B}_{k}}{d z}=2 i \vert \eta \vert \,e^{i \{\phi-2 \lambda_{k}z\}} \hat{B}_{k}^{\dag}, 
\end{equation}
or equally in terms of linear supermodes
\begin{equation} \label{sixB}\nonumber
\frac{d\hat{\mathcal{B}}_{k}}{d z}=i\lambda_{k}\hat{\mathcal{B}}_{k}+2 i \vert \eta \vert \,e^{i \phi} \hat{\mathcal{B}}_{k}^{\dag}. 
\end{equation}
The solution in this basis is exact and given by
\begin{align} \label{k-sol}
\hat{\mathcal{B}}_{k}=\cos(F_{k} z) \hat{\mathcal{B}}_{k}(0) +i\frac{\sin(F_{k} z) }{F_{k}}[\lambda_{k} \hat{\mathcal{B}}_{k}(0)+ 2 \eta \,\hat{\mathcal{B}}_{k}^{\dag}(0)],
\end{align}}
with $F_{k}=\sqrt{\lambda_{k}^{2}-4\vert\eta\vert^{2}}$. For typical evanescent coupling, nonlinearities and pump powers found in quadratic ANWs $\vert \lambda_{k}\vert >2 \vert\eta\vert$ and thus $F_{k} \in \mathbb{R}$. We consider cases only in this power regime in the remainder of the article. Equation (\ref{k-sol})  simplifies into Equations (6)-(7) of ref. \cite{Barral2017b} for the nonlinear directional coupler ($N=2$). The supermodes evolution is similar to the one found there for the individual modes: the power of the SPDC supermode periodically oscillates between a maximum and zero with oscillation periods $L_{k}=\pi/(2 F_{k})$. 

It is interesting to note that waveguide arrays with odd number of waveguides $N$ exhibit a zero supermode $l=(N+1)/2$. As introduced in section \ref{Engineera}, this is a propagation eigenmode with zero eigenvalue $\lambda_{l}=0$ in the slowly-varying amplitude approximation \cite{Efremidis2005}. The oscillation period of the zero-supermode is imaginary $L_{l}=\pi/(4i\vert \eta\vert)$, thus leading to \Da{ the following hyperbolic solution
\begin{align} \label{l-sol}
\hat{\mathcal{B}}_{l}(z)=\cosh(2\vert\eta\vert z) \hat{\mathcal{B}}(0)+i e^{i \phi} \sinh(2\vert\eta\vert z)\hat{\mathcal{B}}_{l}^{\dag}(0).
\end{align}}
Note that Equations (\ref{k-sol}) and (\ref{l-sol}) are respectively the solutions of a non phasematched and perfectly phasematched degenerate parametric amplifiers \cite{Mollow1967}. 
The supermode solution Equation (\ref{k-sol}) can be written in the individual mode basis as the following Bogolyubov transformations
\begin{equation}\label{IndSol1}
\hat{A}_{j}(z) = \sum_{j'=1}^{N} [\tilde{U}_{j, j'}(z) \hat{A}_{j'}(0)+\tilde{V}_{j,j'}(z) \hat{A}_{j'}^{\dag}(0) ],
\end{equation}
where 
\begin{align} \nonumber
\tilde{U}_{j,j'}(z)=\sum_{k=1}^{N} &M_{k,j} M_{k, j'} [\cos(F_{k} z)+i \frac{\lambda_{k}}{F_{k}} \sin(F_{k} z)], \\  \nonumber
\tilde{V}_{j, j'}(z)=\sum_{k=1}^{N} &M_{k,j} M_{k, j'} [\frac{2 i \vert\eta\vert e^{i \phi}} {F_{k}} \sin(F_{k} z)],        
\end{align}
with $\sum_{j=1}^{N}[\vert \tilde{U}_{j, j'}(z) \vert^{2} - \vert \tilde{V}_{j, j'}(z) \vert^{2}]=1$. Note that for $\vert \eta \vert =0$, $\tilde{U}_{j, j'}(z)= U_{j, j'}(z)$ and $\tilde{V}_{j, j'}(z)=0$, with $U_{j, j'}(z)\equiv \sum_{k=1}^{N} M_{k, j} M_{k, j'} e^{i \lambda_{k} z}$, we recover the solution corresponding to the linear array. 

From these equations, it is straightforward to obtain the elements of the covariance matrix $V(z)$, which read 
\begin{align} \nonumber
&V(x_{i}, x_{j})=\sum_{k=1}^{N} \frac{M_{k,i} M_{k,j} }{F_{k}^{2}} \{\lambda_{k}^{2} - 4\vert\eta\vert ^{2} \cos(2F_{k} z) \\  \nonumber
&-4\vert\eta\vert \sin(F_{k} z) [F_{k} \sin(\phi) \cos(F_{k} z) + \lambda_{k} \cos(\phi) \sin(F_{k}z)]\},\\ \nonumber
&V(y_{i}, y_{j})=\sum_{k=1}^{N} \frac{M_{k,i} M_{k,j} }{F_{k}^{2}} \{\lambda_{k}^{2} - 4\vert\eta\vert ^{2} \cos(2F_{k} z) \\  \nonumber
&+4\vert\eta\vert \sin(F_{k} z) [F_{k} \sin(\phi) \cos(F_{k} z) + \lambda_{k} \cos(\phi) \sin(F_{k}z)]\},\\ \nonumber
&V(x_{i}, y_{j})=\sum_{k=1}^{N} \frac{M_{k,i} M_{k,j} }{F_{k}^{2}} \times \\ \label{V0}
&\quad 4\vert\eta\vert \sin(F_{k} z) [F_{k} \cos(\phi) \cos(F_{k} z) - \lambda_{k} \sin(\phi) \sin(F_{k}z)].
\end{align}
This configuration generates quantum correlations between the individual modes --off-diagonal components of the covariance matrix (as shown in Figure \ref{F2}a)--, and hence entanglement is possible in that basis. Likewise, the mean number of signal photons generated in the $j$th waveguide at any propagation length can be directly calculated from Equations (\ref{V0}) as $\bar{N}_{j}=V(x_{j}, x_{j})+V(y_{j}, y_{j})-2$.

Remarkably, the results displayed in this section are general for any ANWs --any evanescent coupling profile $\vec{f}$-- since they are based only on the orthonormality of the supermodes. Equations (\ref{V0}) remain valid for any number of waveguides $N$ or propagation distance $z$. Thus they are a valuable tool which we use in section \ref{Clusters} to engineer linear clusters.

\subsubsection{Flat pump profile: alternating $\pi$ phase}\label{Engineerb2}

When all waveguides are equally pumped such that $\vert\eta_{j}\vert=\vert\eta\vert=\tilde{\eta}$ with an alternating phase $\phi_{j}=(j+1) \pi + \phi$, the joint-spatial supermode matrix Equation (\ref{six}) is notably simplified to 
\begin{equation}\label{L2}\nonumber
\mathcal{L}_{k,k'} (z)=2i \, \delta_{k,N+1-k'} e^{i\{\phi-(\lambda_{k}+\lambda_{k'})\}z}
\end{equation}
via Equation (\ref{Rel1b}). \Da{This pump configuration antidiagonalizes the momentum in the slowly-varying linear supermode basis and the following Heisenberg equations are obtained
\begin{equation} \label{sixC2}\nonumber
\frac{d\hat{B}_{k}}{d z}=2 i \vert \eta \vert e^{i \phi}  \hat{B}_{N+1-k}^{\dag},
\end{equation}
with downconversion gains proportional to $2\vert\eta\vert$.  The above equation can be rewritten in terms of linear supermodes as
\begin{equation}\nonumber
\frac{d\hat{\mathcal{B}}_{k}}{d z}=i\lambda_{k}\hat{\mathcal{B}}_{k}+2 i \vert \eta \vert e^{i \phi}  \hat{\mathcal{B}}_{N+1-k}^{\dag},
\end{equation}
with the following exact solution
\begin{align} \nonumber
\hat{\mathcal{B}}_{k}(z)=[\cosh(2\vert\eta\vert z) \hat{\mathcal{B}}_{k}(0)+i e^{i \phi} &\sinh(2\vert\eta\vert z)\hat{\mathcal{B}}_{N+1-k}^{\dag}(0)] \\ \label{sixD2}
&\qquad \qquad \times e^{i\lambda_{k} z}.
\end{align}}
Note that this is the solution of a perfectly phase-matched nondegenerate parametric amplifier \cite{Mollow1967}. The supermode solution Equation (\ref{sixD2}) can be written in the individual mode basis as the following transformation
\begin{align}\label{IndSol2} \nonumber
\hat{A}_{j}(z&) = \sum_{j'=1}^{N} {U}_{j, j'}(z) \times \\
&[\cosh(2\vert\eta\vert z) \hat{A}_{j'}(0) + (-1)^{j'+1} i e^{i \phi} \sinh(2\vert\eta\vert z) \hat{A}_{j'}^{\dag}(0) ],
\end{align} 
where we have used Equations (\ref{Rel1a}), (\ref{Rel1b}) and the propagator related to the linear array ${U}_{j, j'}(z)$ introduced above. The solution is thus decoupled in this configuration: input single-mode squeezed states of light squeezed along the axis $(j'+1) \pi + \phi $ propagate in the corresponding linear array with propagation matrix ${U}_{j, j'}(z)$. From this equation, after a long but straightforward calculation, we obtain the elements of the covariance matrix $V(z)$, which read as follows
\begin{align} \nonumber
&V(x_{i}, x_{j})=[\cosh(4\vert \eta \vert z) + (-1)^{j} \sin(\phi) \sinh(4\vert \eta \vert z)] \,\delta_{i,j}, \\ \nonumber
&V(y_{i}, y_{j})=[\cosh(4\vert \eta \vert z) - (-1)^{j} \sin(\phi) \sinh(4\vert \eta \vert z)] \,\delta_{i,j}, \\ \label{Vpi}
&V(x_{i}, y_{j})=(-1)^{j} \cos(\phi) \sinh(4\vert \eta \vert z) \,\delta_{i,j}.
\end{align}
Then, in this case quantum correlations are efficiently generated in the supermode basis but they disappear in the individual mode basis --no off-diagonal elements of the covariance matrix (Figure \ref{F2}b)--. The device  thus produces independent squeezed fields. The results obtained in this section are general for any coupling profile $\vec{f}$ since they rely on Equations (\ref{Rel1}) - (\ref{Rel1b}) only.
Equations (\ref{Vpi}) remain valid for any number of waveguides $N$ or propagation distance $z$.  Notably, this is an interesting regime for discrete variables since N-dimensional two-photon NOON states \Da{can be postselected} \cite{Barral2019c}.

\subsubsection{Flat pump profile: any alternating phase} \label{Engineerb3}

Both cases analyzed in sections \ref{Engineer}.B.1 and \ref{Engineer}.B.2 are encompassed through the use of Equations (\ref{Rel1c}) - (\ref{Rel1d}). In the case of an array composed of N waveguides equally pumped such that $\vert\eta_{j}\vert=\vert\eta\vert=\tilde{\eta}$ and alternating phases $\phi_{2j}$ and $\phi_{2j-1}$, the joint-spatial supermode matrix Equation (\ref{six}) is notably simplified to 
\begin{align}\nonumber
&\mathcal{L}_{k,k'} (z)=\\ \label{Case3}
&2i e^{i\Delta\phi^{+}}[\cos{(\Delta\phi^{-})} e^{-2i\lambda_{k} z}  \delta_{k,k'} - i \sin{(\Delta\phi^{-})} \delta_{k,N+1-k'} ],
\end{align}
with $\Delta\phi^{\pm}=(\phi_{2j}\pm\phi_{2j-1})/2$. Thus the solution of the system oscillates between Equations (\ref{L1}) and (\ref{L2}) for a general phase difference $\Delta\phi^{-}$. Particularly, for $\phi_{2j}=\phi+\pi/2$ and $\phi_{2j-1}=\phi$ both the diagonal and antidiagonal terms have the same weight such as
\begin{equation}\nonumber
\mathcal{L}_{k,k'} (z)=\sqrt{2} \,i e^{i(\phi + \pi/4)}\,[e^{-2i \lambda_{k} z} \delta_{k,k'} -i \,\delta_{k,N+1-k'}].
\end{equation}
The solution will present then both oscillatory and hyperbolic terms. More light is shed on the features that this configuration produces in section \ref{Multimode}.

\subsubsection{Pumping only the even or odd waveguides} \label{Engineerb4}

Another simplified joint-spatial supermode matrix is obtained if either even waveguides only ($\vert\eta_{2j}\vert=\tilde{\eta}$, $\vert\eta_{2j-1}\vert=0$ and $\phi_{2j}=\phi$) or odd waveguides only ($\vert \eta_{2j-1}\vert=\tilde{\eta}$, $\vert\eta_{2j}\vert=0$ and $\phi_{2j-1}=\phi$) are pumped, such that
\begin{equation}\nonumber
\mathcal{L}_{k,k'} (z)=i e^{i\phi}\,[e^{-2i \lambda_{k} z} \delta_{k,k'} \pm \delta_{k,N+1-k'}],
\end{equation}
with plus for odd and minus for an even pump profile through Equations (\ref{Rel1c}) - (\ref{Rel1d}). The solutions are here more complex that those of cases \ref{Engineerb1} and \ref{Engineerb2}. \Da{For instance, pumping only the odd waveguides we have the following Heisenberg equation for the slowly-varying linear supermodes
\begin{equation} \nonumber
\frac{d\hat{B}_{k}}{d z}=i \vert \eta \vert e^{i \phi}  (e^{-2i\lambda_{k}z} \hat{B}_{k}^{\dag}+\hat{B}_{N+1-k}^{\dag}).
\end{equation}
Equally, for the linear supermodes we get
\begin{equation} \nonumber
\frac{d\hat{\mathcal{B}}_{k}}{d z}=i\lambda_{k}\hat{\mathcal{B}}_{k}+i \vert \eta \vert e^{i \phi}  (\hat{\mathcal{B}}_{k}^{\dag}+\hat{\mathcal{B}}_{N+1-k}^{\dag}),
\end{equation}
with the following exact solution
\begin{align}\nonumber
&\hat{\mathcal{B}}_{k}(z)=\{\cosh(\vert\eta\vert z) [\cos(\tilde{F}_{k} z)+i\frac{\lambda_{k}}{\tilde{F}_{k}}\sin(\tilde{F}_{k} z) ]\hat{\mathcal{B}}_{k}(0)\\  \nonumber
&+i\frac{\vert \eta \vert}{\tilde{F}_{k}}\sin(\tilde{F}_{k} z) [\cosh(\vert\eta\vert z) \hat{\mathcal{B}}_{k}^{\dag}(0) -i\sinh(\vert\eta\vert z)  \hat{\mathcal{B}}_{N+1-k}(0)]\\ \label{BEO}
&+i\sinh(\vert\eta\vert z) [\cos(\tilde{F}_{k} z)+i\frac{\lambda_{k}}{\tilde{F}_{k}}\sin(\tilde{F}_{k} z) ]\hat{\mathcal{B}}_{N+1-k}^{\dag}(0)\},
\end{align}}
with $\tilde{F}_{k}=\sqrt{\lambda_{k}^{2}-\vert\eta\vert^{2}}$ and where we have set $\phi=0$ for the sake of simplicity. This solution shows that the side linear supermodes are symmetrically coupled two by two --$k$ with $N+1-k$--, being the zero supermode the only one independently squeezed. The solution in the individual mode basis can be written as Equation (\ref{IndSol1}) with
\begin{align} \nonumber
\tilde{U}_{j,j'}(z)=\sum_{k=1}^{N} M_{k, j} M_{k, j'} [\cosh(\vert\eta\vert z)&\cos(\tilde{F}_{k} z) \\ \nonumber
+i \frac{\lambda_{k}\cosh(\vert\eta\vert z)+(-1)^{j'+1}\vert \eta \vert \sinh(\vert\eta\vert z)}{\tilde{F}_{k}} &\sin(\tilde{F}_{k} z)], \\ \nonumber
\tilde{V}_{j, j'}(z)=\sum_{k=1}^{N} M_{k, j} M_{k, j'} (-1)^{j'+1} [i\sinh(\vert\eta\vert z)&\cos(\tilde{F}_{k} z) \\   \nonumber
+ i\frac{i\lambda_{k}\sinh(\vert\eta\vert z)+(-1)^{j'+1} \vert \eta \vert \cosh(\vert\eta\vert z)}{\tilde{F}_{k}} &\sin(\tilde{F}_{k} z)],\end{align}
where we have used the property of the supermodes Equation (\ref{Rel1a}). The elements of the covariance matrix $V(z)$ are in this case the following
\begin{align} \nonumber
V(x_{i}, x_{j})=\sum_{k=1}^{N} & M_{k, i} M_{k, j} [\cosh(2\vert\eta\vert z) (\frac{\lambda_{k}+\vert \eta \vert \cos(2\tilde{F}_{k}z)}{\lambda_{k}+\vert \eta \vert})\\ \nonumber
&+(-1)^{j}\sqrt{\frac{\lambda_{k}-\vert \eta \vert}{\lambda_{k}+\vert \eta \vert}} \sinh(2\vert\eta\vert z)\sin(2\tilde{F}_{k}z)],
\\  \nonumber
V(y_{i}, y_{j})=\sum_{k=1}^{N} & M_{k, i} M_{k, j} [\cosh(2\vert\eta\vert z) (\frac{\lambda_{k}-\vert \eta \vert \cos(2\tilde{F}_{k}z)}{\lambda_{k}-\vert \eta \vert})\\ \nonumber
&-(-1)^{j}\sqrt{\frac{\lambda_{k}+\vert \eta \vert}{\lambda_{k}-\vert \eta \vert}} \sinh(2\vert\eta\vert z)\sin(2\tilde{F}_{k}z)],
\\ \nonumber
V(x_{i}, y_{j})=\sum_{k=1}^{N} & M_{k, i} M_{k, j} [(-1)^{j+1} \sinh(2\vert\eta\vert z)\cos(2\tilde{F}_{k}z) \\ \nonumber
&\qquad \qquad + \frac{\vert \eta \vert}{\tilde{F}_{k}} \cosh(2\vert\eta\vert z)\sin(2\tilde{F}_{k}z) ].
\end{align}
Thus, quantum correlations between the individual modes are generated. These solutions generalize to $N$ dimensions the paradigmatic example of pumping one waveguide in a nonlinear directional coupler \cite{Kruse2015}. Similar solutions are obtained when pumping only the even waveguides.

\subsubsection{Pumping the central waveguide in an odd ANW}  \label{Engineerb5}

A common and simple way of pumping an odd ANWs is to inject the pump only in the central waveguide $j=l\equiv(N+1)/2$ (see Figure \ref{F1}a) \cite{Solntsev2014}. The following joint-spatial supermode distribution is then obtained
\begin{equation}\label{onepump}
\mathcal{L}_{k,k'} (z)=2 i e^{i\phi_{l}} M_{k, l} M_{k', l} \,e^{-i(\lambda_{k}+\lambda_{k'}) z}.
\end{equation}
Notably, in the case of symmetric coupling profile arrays like the homogeneous or the parabolic profiles shown above, the elements of the zero supermode have zeros in the even elements, i.e. $M_{k,l}=0$ for $k$ even. Thus, only odd supermodes are produced in the ANWs under this configuration. For instance, for N=5 and a homogeneous coupling profile we obtain as approximated solutions Equation (\ref{l-sol}) for the zero supermode ($l=3$) and Equation (\ref{sixD2}) for the $k=1,5$ side supermodes after rescaling $\vert\eta\vert$ to $\vert\eta\vert/l$. Figure \ref{F2}c shows the covariance matrix in the individual mode basis related to this pump configuration in an ANWs with a homogeneous coupling profile. 

\vspace{0.5cm}
The above five cases exhibit the versatility of the ANWs through pump engineering and shed light on propagation in these devices. We further discuss the relationship between linear and nonlinear supermodes, and the generated squeezing along propagation in section \ref{Multimode}. The generation of multipartite entanglement in ANW has been recently tackled in \cite{Barral2020, Barral2020b} and we further present in section \ref{Clusters} an efficient protocol for the generation of linear cluster states based on the analytical solution obtained in section \ref{Engineerb1}.


\subsection{Phase-matching engineering}\label{Engineerc}

A common phase-matching technique for efficient frequency conversion in $\chi^{(2)}$ nonlinear waveguides is obtained through wavevector quasi-phase matching ($\Delta \beta$-QPM). A standard implementation of $\Delta \beta$-QPM is periodical inversion of the second-order susceptibility $\chi^{(2)}$ with period $\Lambda_{\Delta\beta}=2\pi/\Delta\beta$, like for instance in PPLN waveguides \cite{Alibart2016}. However, in the case of waveguide arrays, a second cause of phase mismatch --the coupling-- is present, as shown in Equation (\ref{six}). In this case a similar strategy can be used to phase match specific supermodes through a second periodical inversion $\Lambda_{C}(k')$ --coupling quasi-phase matching (C-QPM)-- \cite{Barral2019c}. This slow modulation will match the propagation constant $\lambda_{k'}$ of the $k'$th slowly varying supermode amplitude. We consider, for instance, a homogeneous coupling profile where $\lambda_{k'}=-\lambda_{N+1-k'}\equiv 2C_{0} \cos{[k'\pi/(N+1)]}$. In this case, the periodical inversion --coupling period-- can be set as $\Lambda_{C}({k'})=\left| \pi/\lambda_{k'}\right|$, thus phase matching the $k'$th and $(N+1-k')$th side supermodes. Equation (\ref{six}) is then written as
\begin{align} \nonumber
&\mathcal{L}_{k,k'} (z)\approx  \\ \label{Lqpm}
&\frac{8 i}{\pi} \sum_{j=1}^{N} \frac{\vert \eta_{j}\vert}{\tilde{\eta}} M_{k,j} M_{k',j} \cos{(2\lambda_{k'} z)} e^{i\{\phi_{j}-(\lambda_{k}+\lambda_{k'})z\}},
\end{align}
where we have used the first-order Fourier series of square-wave C-QPM domains with duty cycles of $50\%$. Thus, using a flat pump profile, Equation (\ref{Lqpm}) is simplified to 
\begin{equation}\nonumber
\mathcal{L}_{k,k'} (z)\approx \frac{8i}{\pi} \cos{(2\lambda_{k'} z)} \delta_{k,k'} \,e^{i \{\phi-(\lambda_{k}+\lambda_{k'})z\}},
\end{equation}
and the Heisenberg equations read
\begin{align} \nonumber
\frac{d\hat{B}_{k}}{d z} &\approx \frac{4 i \tilde{\eta}}{\pi} e^{i \phi} \hat{B}_{k}^{\dag},  \quad &k=k', N+1-k',\\   \nonumber
\frac{d\hat{B}_{k}}{d z} &\approx  \frac{4 i \tilde{\eta}}{\pi} e^{i \{\phi-2 \lambda_{k}z\}} \hat{B}_{k}^{\dag}, \quad &k\neq k', N+1-k'.
\end{align}
Hyperbolic solutions as Equation (\ref{l-sol}) are obtained for the $k'$th and $(N+1-k')$th supermodes and oscillatory solutions like Equation (\ref{k-sol}) for the other supermodes. Note that the gains are reduced by a factor $2/\pi$ in comparison with the no C-QPM case which can be compensated with a propagation distance $\pi/2$ longer. 

This powerful technique allows to control the supermodes efficiently building up. In terms of individual-modes entanglement it would be interesting to build up supermodes but with light in all the individual modes.  Remarkably, in the case of parabolic arrays with an even number of waveguides all the supermodes can efficiently build up. This interesting case will be presented elsewhere.
\begin{figure}[t]
  \centering
    \includegraphics[width=0.48\textwidth]{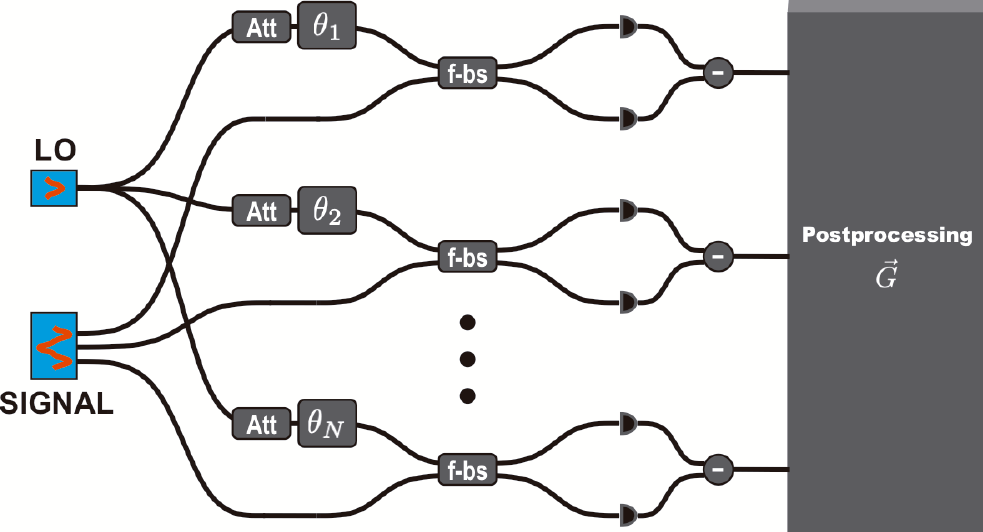}
\vspace {0cm}\,
\hspace{0cm}\caption{\label{F1b}\small{Fibered multimode balanced homodyne detector. The SPDC signal modes are collected into optical fibers by a V-groove array and directed to a multimode balanced homodyne detector (BHD) where modes are measured using adapted LO phase profile $\vec{\theta}$, electronic gain profile $\vec{G}$ and suitable postprocessing --addition and subtraction-- of electronic signals. {Att} and {f-bs} stand respectively for attenuator and 3 dB fiber beam splitter.}}
\end{figure}
\begin{figure*}
\centering
    {\raisebox{0.4cm}{\begin{turn}{90} Flat pump - Uniform phase \end{turn} \hspace{0.2cm}  }}\subfigure{\includegraphics[width=0.3\textwidth]{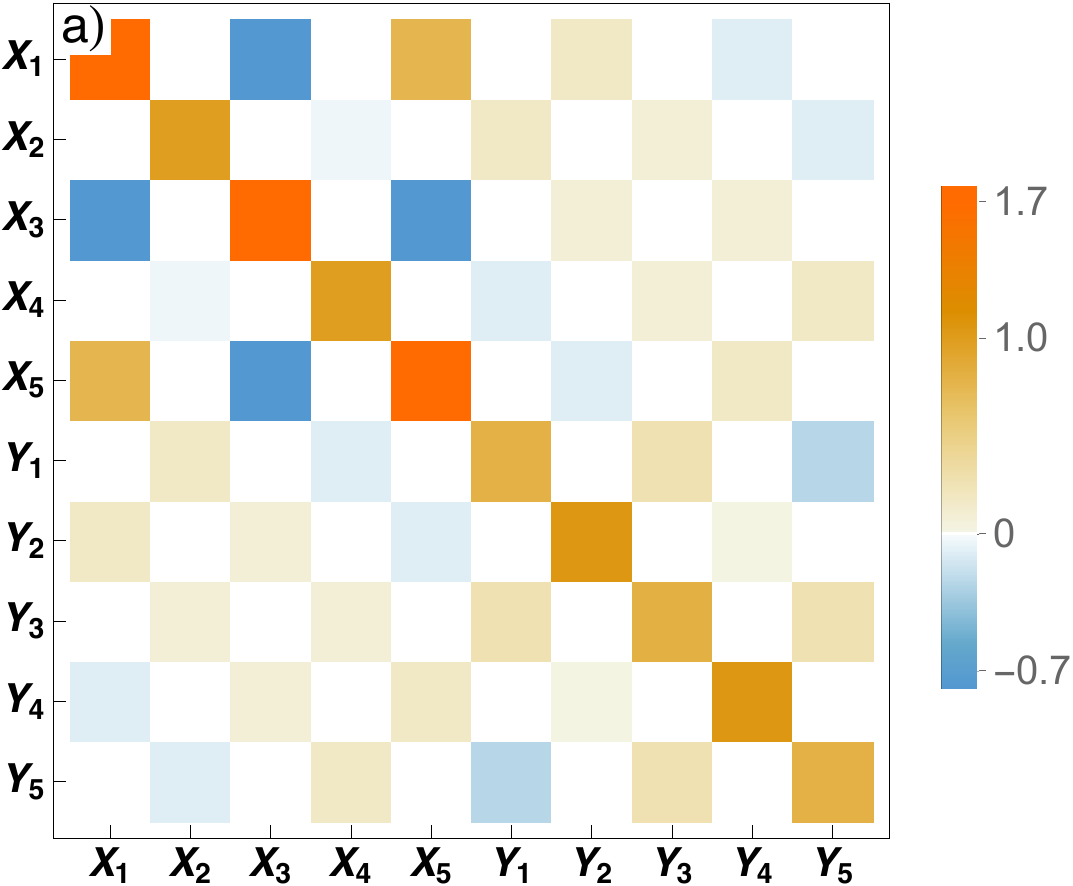}} 
    \vspace {0cm}
    \quad \subfigure{\includegraphics[width=0.3\textwidth]{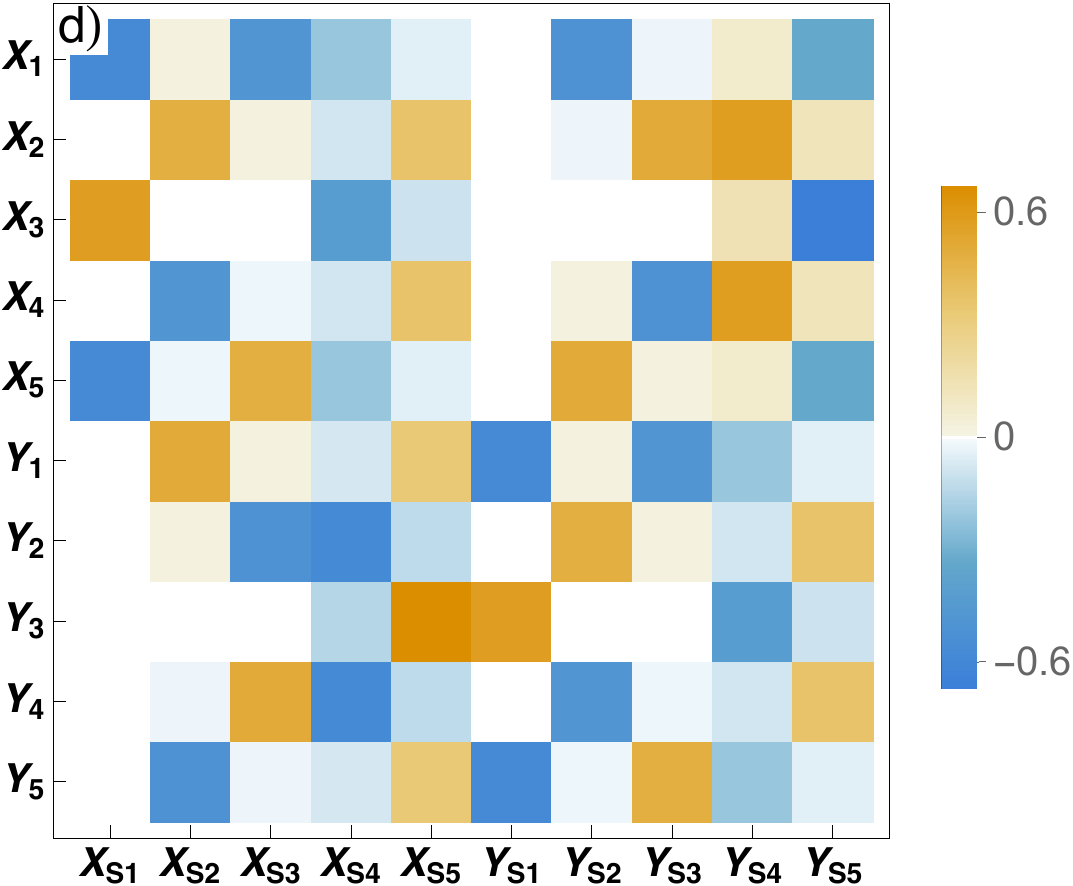}}
     \quad \subfigure{\includegraphics[width=0.3\textwidth]{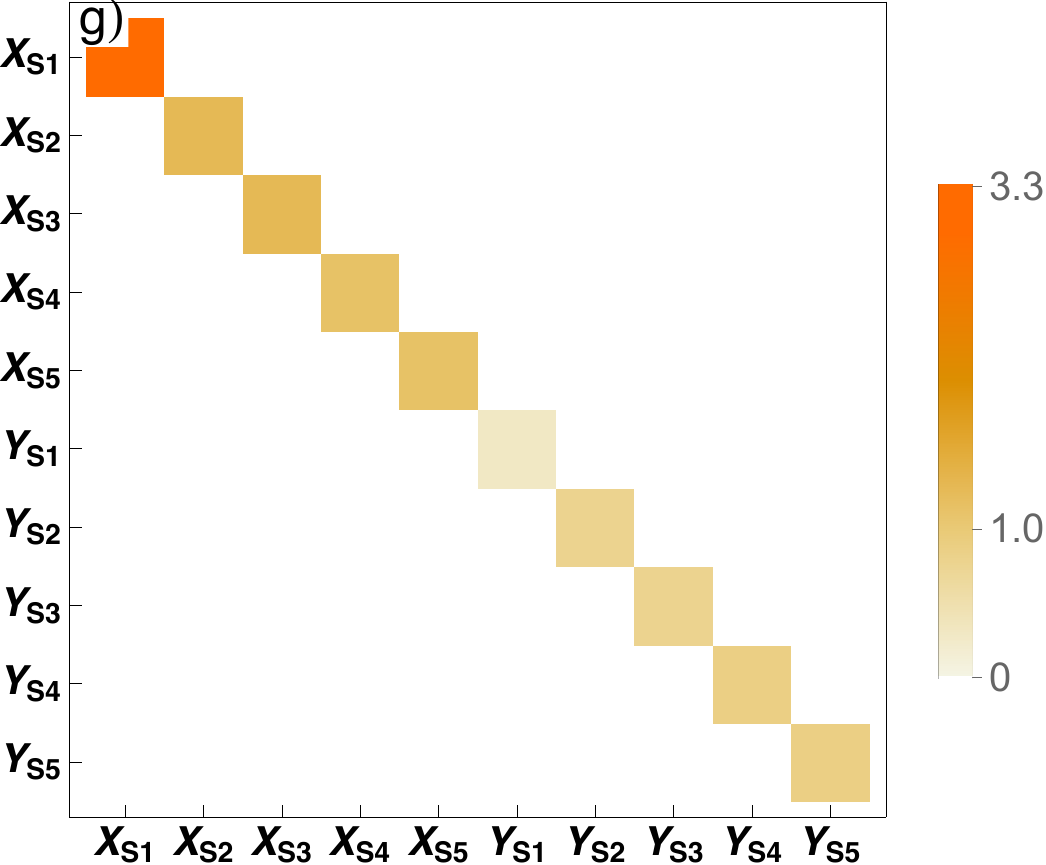}} \\
    
        {\raisebox{0cm}{\begin{turn}{90} Flat pump - Alternating $\pi$ phase \end{turn} \hspace{0.2cm}  }}\subfigure{\includegraphics[width=0.3\textwidth]{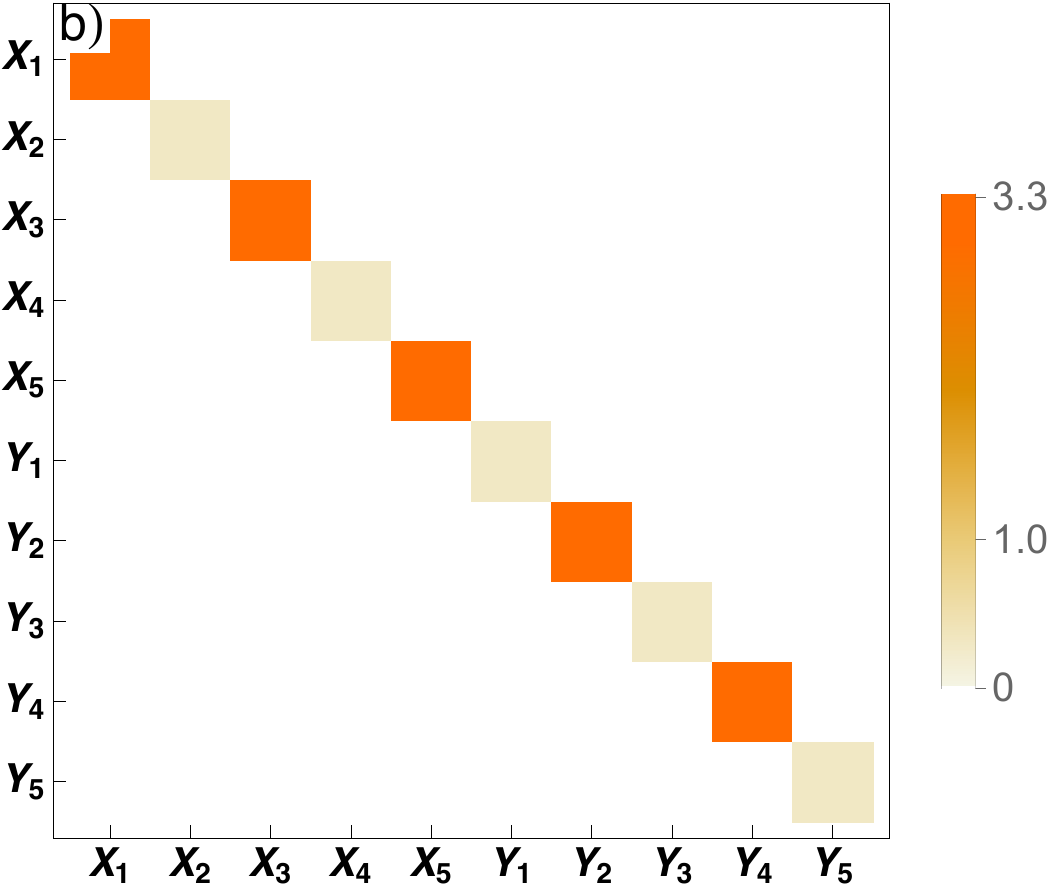}} 
    \vspace {0cm}
     \quad \subfigure{\includegraphics[width=0.3\textwidth]{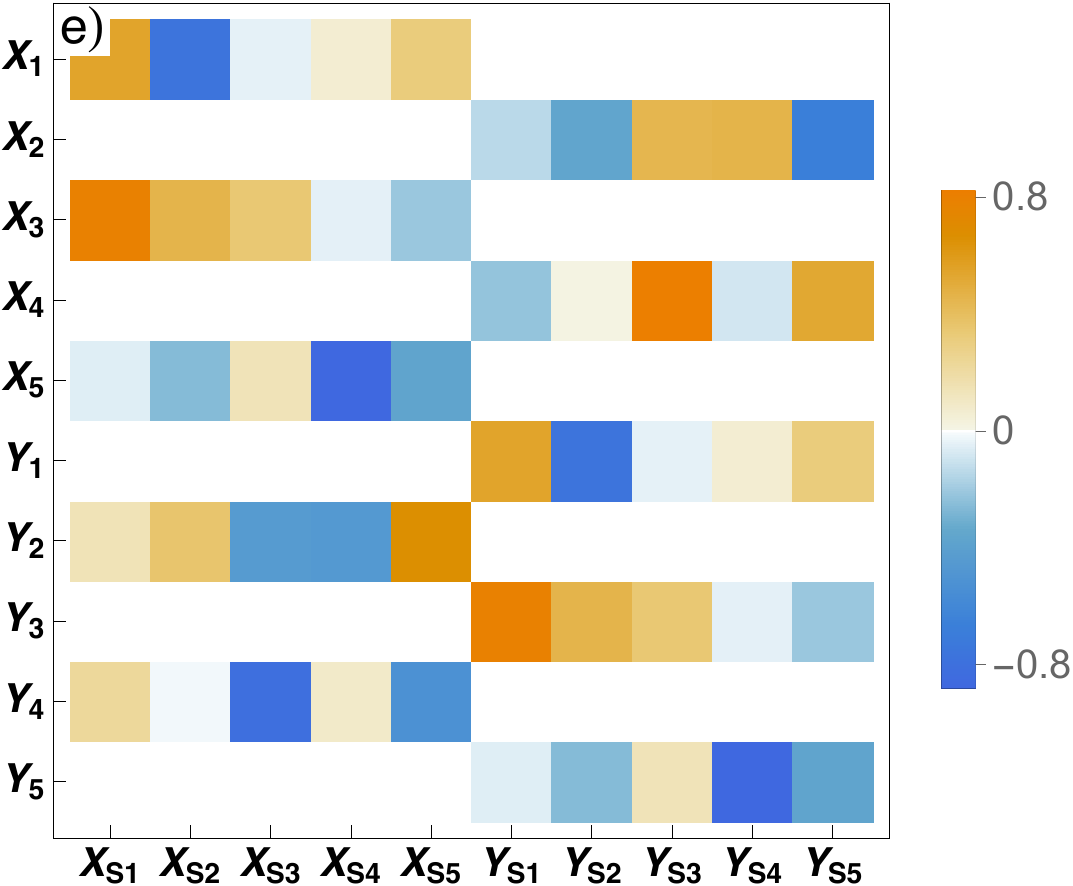}}
        \vspace {0cm}
     \quad \subfigure{\includegraphics[width=0.3\textwidth]{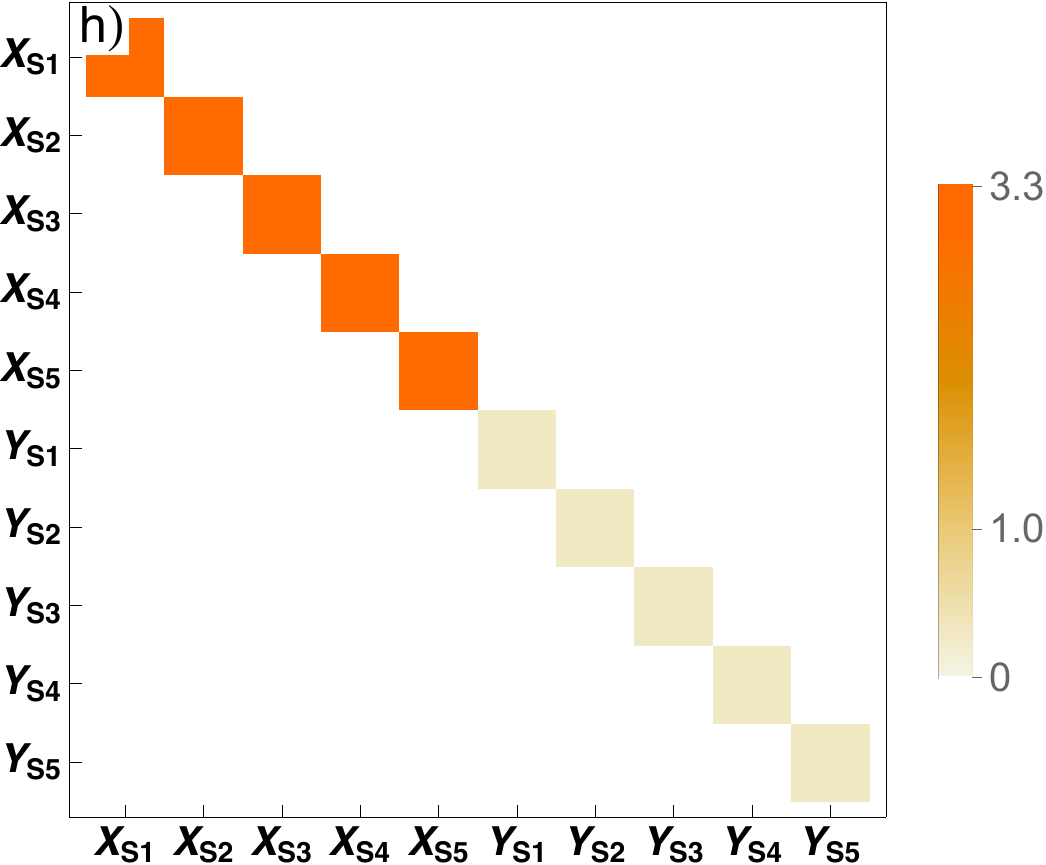}} \\
     
    {\raisebox{0.7cm}{\begin{turn}{90} Pump central waveguide \end{turn} \hspace{0.2cm}  }}\stackunder[5pt]{\subfigure{\includegraphics[width=0.3\textwidth]{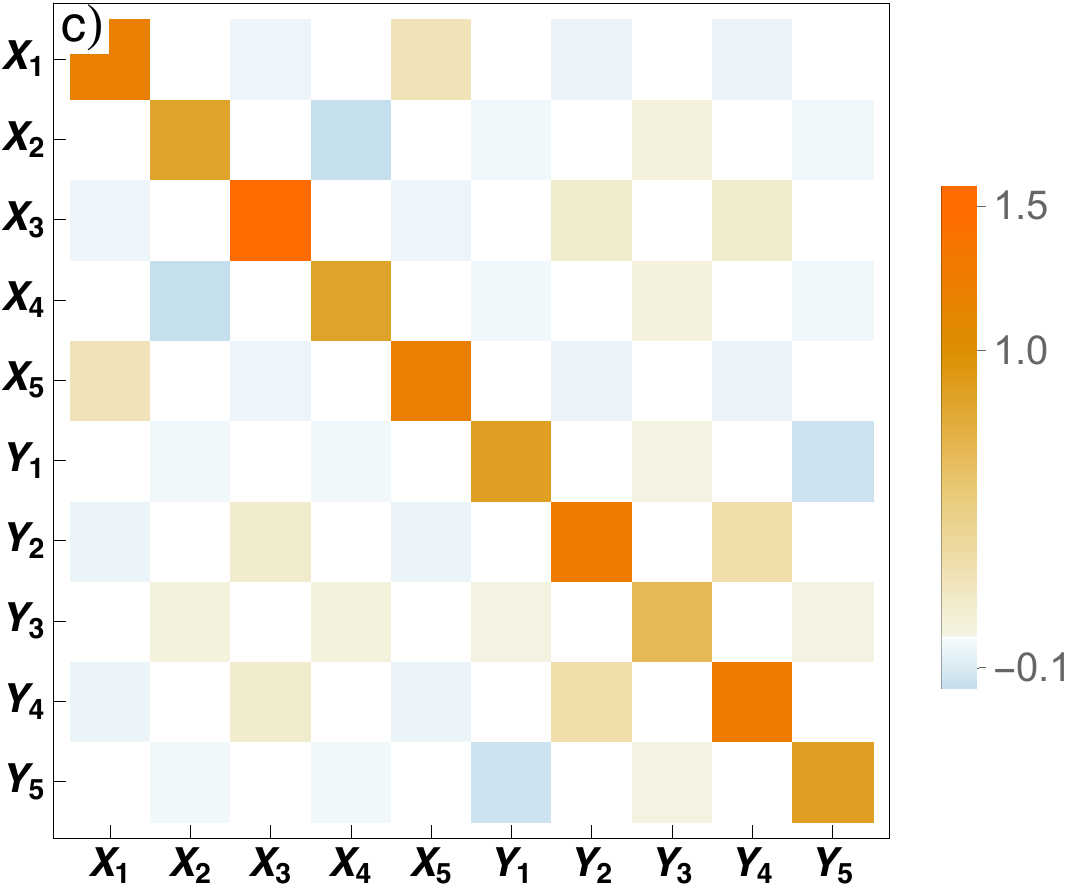}}}  {$V(z)$} 
    \vspace {0cm}
     \quad  \stackunder[5pt]{\subfigure{\includegraphics[width=0.3\textwidth]{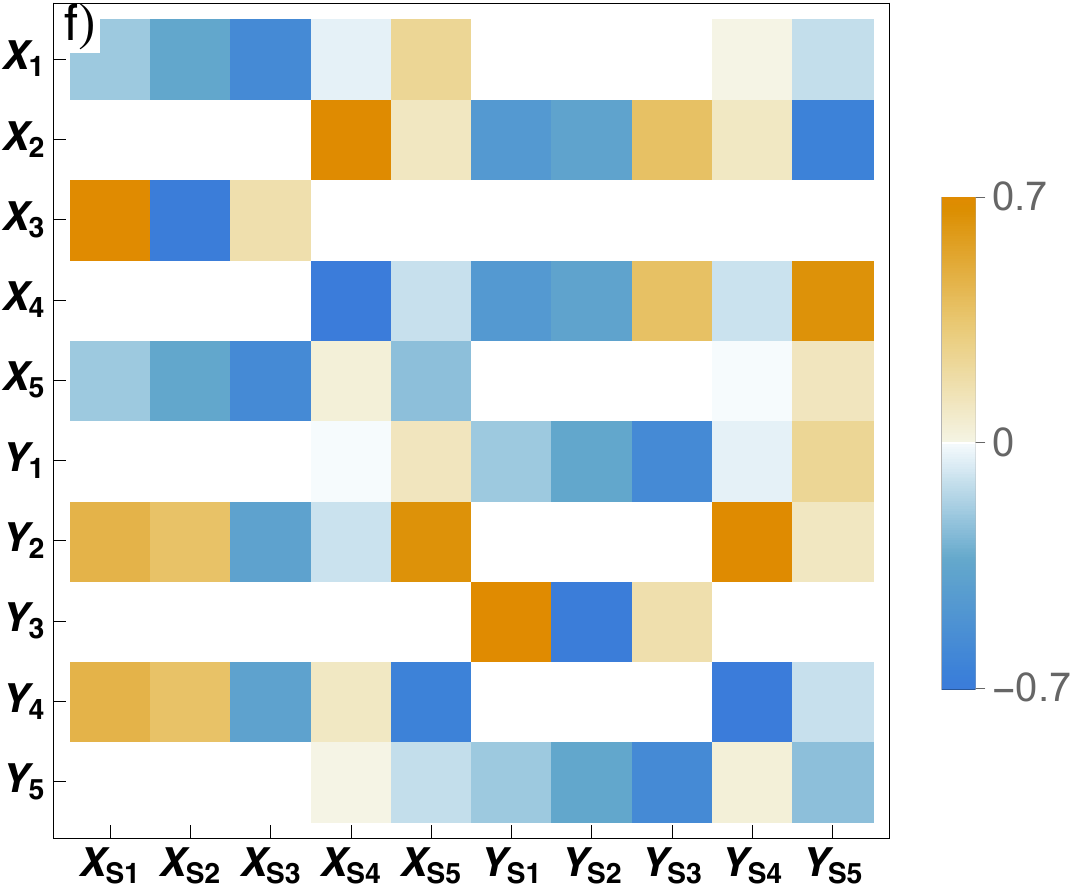}}}{$R_{1}(z)$}
        \vspace {0cm}
    \quad  \stackunder[5pt]{\subfigure{\includegraphics[width=0.3\textwidth]{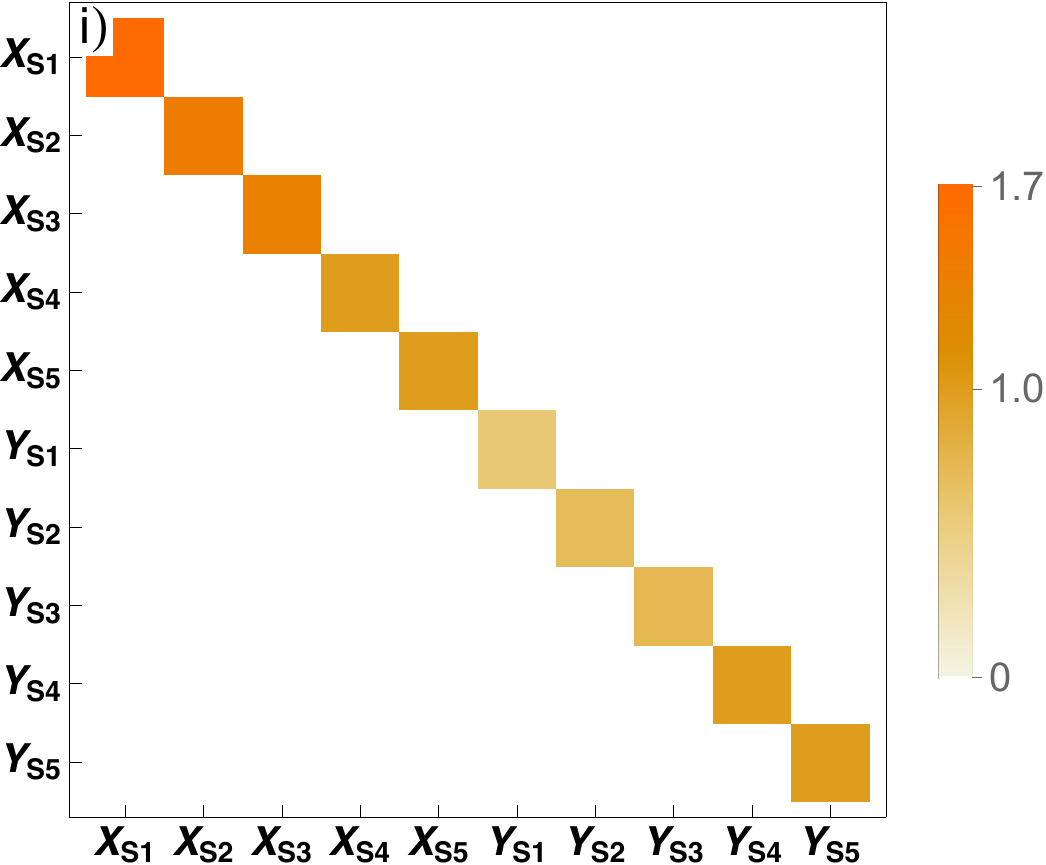}}} {$K^{2}(z)$} \\
    
\vspace {0cm}
\hspace{0cm}\caption{\label{F2}\small{Covariance matrices in the individual mode basis $V(z)$ (a-c), Bloch-Messiah's transformation matrices $R_{1}(z)$ (d-f), and diagonal covariance matrices in the nonlinear supermode mode basis $K^{2}(z)$ (g-i) for a five-waveguides homogeneous coupling-profile ANWs. The upper row displays the results obtained for a flat pump profile with a uniform phase ($\Delta\phi^-=0$): Equation (\ref{V0}) with $\phi=-\pi/2$. The central row displays the results obtained for a flat pump profile with an alternative $\pi$ phase ($\Delta\phi^-=\pi$): Equation (\ref{Vpi}) with $\phi=-\pi/2$. The lower row displays the results obtained pumping only the central waveguide. We applied Equation (\ref{onepump}) into Equation (\ref{Hei}) and solved numerically for $\phi_{l}=-\pi/2$. We set typical parameters in PPLN waveguides: $C_{0}=0.24$ mm$^{-1}$, $\eta=0.015$ mm$^{-1}$ and $z=20$ mm. Absolute values lower than $10^{-2}$ are shown in white for the sake of exposition.
}}
\end{figure*}

\subsection{Balanced homodyne detection}\label{Engineerd}

The measurement of quantum noise variances and correlations is carried out by multimode balanced homodyne detection (BHD) \cite{Fabre2019}. In a fully fibered approach the multimode squeezed state generated in the array can be collected in optical fibers through a V-groove array. A laser at signal frequency is demultiplexed into a number of individual optical fibers with fiber attenuators and phase shifters and individually mixed with the output SPDC through 3 dB fibered beam splitters as sketched in Figure \ref{F1b}. Each pair of mixed signals is sent to a BHD where the current of each photodiode is subtracted and suitably amplified. 

We point out that the spatial profile of the LO in the multimode BHD has to be adapted to the strategy of entanglement generation and the entangled state basis. Access to the quantum information encoded in the individual or any of the supermode bases will indeed depend on a suitable BHD \cite{Dauria2009}. Figure \ref{F1b} displays the possible knobs at the measurement stage. The local oscillator can be tuned to detect correlations between the individual output signals of the array -- in the individual mode basis $\hat{\mathcal{A}}$ -- or shaped to match any supermode -- in the linear supermode basis $\hat{\mathcal{B}}$, nonlinear supermode basis $\hat{\mathcal{C}}$ or any other basis. This LO shaping can be accomplished using attenuator and phase shifters or can be emulated with electronic gains. Remarkably, LO shaping enables the measurement of entangled states encoded in bases based on nonlinear supermodes. We have discussed these issues elsewhere \cite{Barral2020} and mentioned the existing implementations in other domains \cite{Armstrong2012, Cai2017}.

\section{Multimode squeezing}\label{Multimode}

The ANW is a natural platform for generating multimode squeezing due to the distributed coupling and nonlinearity. Such a distributed configuration combines the necessary squeezing and coupling instrumental to produce multimode entanglement in a new way, accessible only to guided-wave nonlinear components. As shown above, the different methods of detection available enable the encoding of quantum information in three ways: individual modes, linear and nonlinear supermodes. We now investigate multimode squeezing focusing on the nonlinear supermodes, which maximize the resources available at every propagation length in terms of squeezing and can generate entanglement through LO shaping or emulation \cite{Barral2020}. We thus display in this section the outcomes in terms of {squeezing obtained from numerical solutions of Equation (\ref{BMD}) using the general method of Bloch-Messiah. The large parameter space of the ANW enables an infinite number of configurations. We focus on the configurations displayed in section \ref{Engineer} in terms of pumping profile --in amplitude and phase--, propagation length  and coupling profile. This allows us to discuss the relationship of these numerical results in terms of nonlinear supermodes with the analytical solutions obtained in section \ref{Engineer} in terms of individual modes and linear supermodes.}

{For the purpose of assessing multimode squeezing, i) we start displaying the connection between the individual and nonlinear supermode bases through the covariance matrices in both bases at a fixed propagation length. We then move to the evolution of squeezing in the nonlinear supermode basis. ii) We first focus on the case of flat pumping to ii.a) analyze the evolution of squeezing along propagation with any alternating pumping phase, and to ii.b) display the influence of different coupling profiles and strengths on the squeezing. Finally, we discuss ii.c) the connection between the nonlinear supermodes and the linear supermodes for an uniform phase.  iii) We finish exhibiting the squeezing obtained for other pumping profiles: iii.a) pumping only the odd waveguides, and iii.b) the simplest configuration, pumping only the central waveguide in an ANW with an odd number of waveguides.}


\subsubsection{Covariance matrices in the individual and nonlinear supermode bases}

We begin with the analysis of Figure \ref{F2} {where we compare the covariance matrices in the individual and nonlinear supermode bases at a propagation distance of $z=20$ mm. We exhibit the covariance matrices in the individual mode basis $V(z)$ in Figures \ref{F2}a-\ref{F2}c, the Bloch-Messiah's transformation matrices $R_{1}(z)$ in Figures \ref{F2}d-\ref{F2}f, and the diagonal covariance matrices in the nonlinear supermode mode basis $K^{2}(z)$ in Figures \ref{F2}g-\ref{F2}i. The Bloch-Messiah decomposition of Equation (\ref{BMD}) has been computed numerically.} We display three of the pumping cases analyzed in Section \ref{Engineer} in a 5-waveguide homogeneous coupling-profile ANW. Figures \ref{F2}a and \ref{F2}b display respectively the results for a flat pump profile and uniform phase [$\Delta \phi^-=0$, Equation (\ref{V0})], where strong quantum correlations between specific quadratures of the fields are generated, and for a flat pump profile and alternating $\pi$ phase [$\Delta \phi^-=\pi$, Equation (\ref{Vpi})], where single-mode squeezing is generated but no correlation in achieved in the individual basis. Figures \ref{F2}g and \ref{F2}h display the respective diagonal covariance matrix in the supermode basis, and Figures \ref{F2}d and \ref{F2}e the corresponding transformations between individual and supermode bases. Likewise, Figure \ref{F2}c displays the covariance matrix in the individual mode basis obtained when pumping only the central waveguide. This case resembles the one shown in Figure \ref{F2}a, with a similar topology of quantum correlations, but different strength and sign. Figures \ref{F2}i and \ref{F2}f display the diagonal covariance matrix $K^{2}(z)$ and the transformation between bases $R_{1}(z)$, making more obvious the difference with the flat pump case of Figures \ref{F2}d and \ref{F2}g. Overall, these figures show the versatility of our approach, yielding different multimode squeezing features for different input pump profiles. We analyze in more depth the obtained squeezing $K^{2}_{N+m}$ exploring further the parameter space along propagation $z$, coupling $C_{j}$ and pumping $\eta_{j}$.
\begin{figure}[t]
  \centering
    \subfigure{\includegraphics[width=0.445\textwidth]{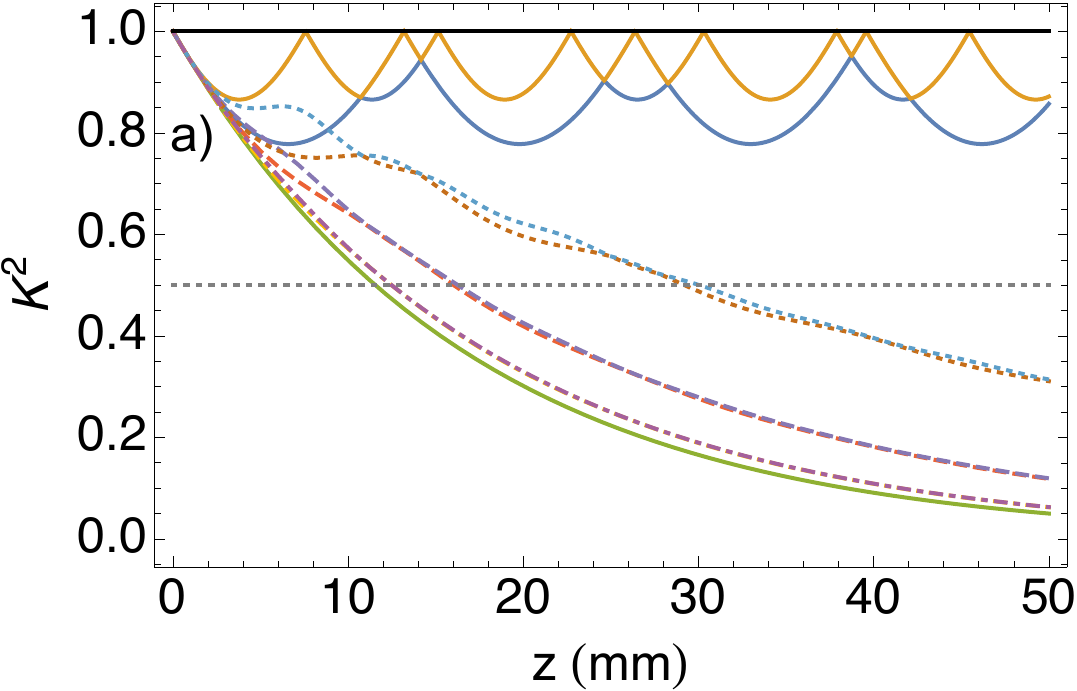}}
    \vspace {0cm}\,
    \subfigure{\includegraphics[width=0.445\textwidth]{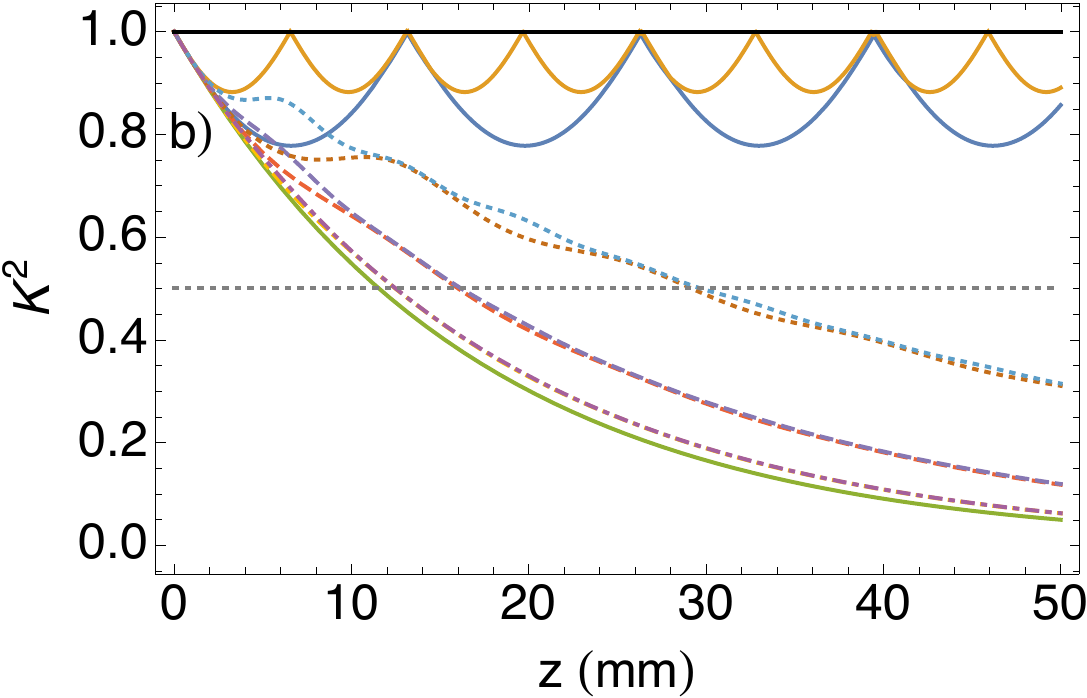}}
    \vspace {0cm}\,
    \subfigure{\includegraphics[width=0.445\textwidth]{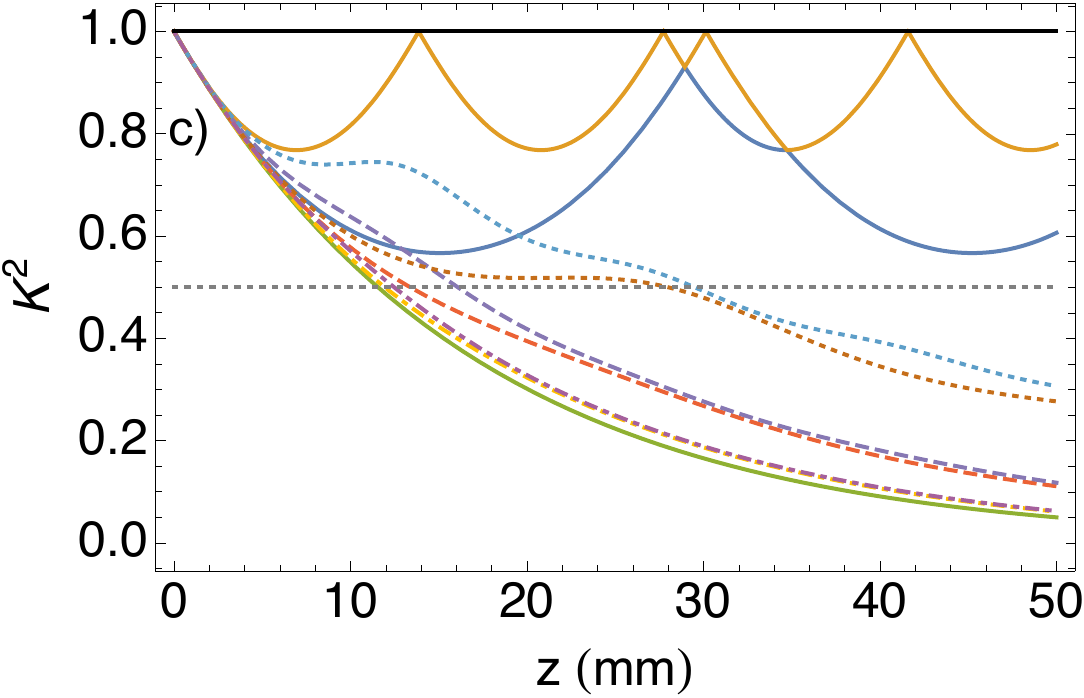}}
\vspace {0cm}
\hspace{0cm}\caption{\label{F3}\small{Evolution of nonlinear supermode squeezing $K^{2}_{N+m}(z)$ in a five-waveguides (a) homogeneous, (b) parabolic and (c) square-root coupling-profile nonlinear array. The zero supermode (\David{$m=l=3$}) is always efficiently squeezed independently of $\Delta\phi^{-}$ (lower solid curve, green). $\Delta\phi^{-}=0$ (upper solid curves, blue and orange), $\Delta\phi^{-}=\pi/2$ (solid, green) and intermediate cases: $\Delta\phi^{-}=\pi/8$ (dotted), $\Delta\phi^{-}=\pi/4$ (dashed) and $\Delta\phi^{-}=3\pi/8$ (dot-dashed). 3 dB squeezing level in dotted gray. $C_{0}=0.24$ mm$^{-1}$ for (a)-(b). $C_{0}=0.08$ mm$^{-1}$ for (c). $\eta=0.015$ mm$^{-1}$.}}
\end{figure}

\subsubsection{Evolution of multimode squeezing for a flat pump profile}

{\it a. Alternating pumping phase $\Delta\phi^{-}$ and nonlinear supermodes squeezing behavior}. Figure \ref{F3} shows the evolution of noise squeezing ($K^{2}_{N+m}(z)<1$) of the five nonlinear supermodes for a flat pump profile in a $N=5$ ANW. We show the effect of the coupling profile $\vec{f}$, the value of the coupling constant $C_{0}$ and the relative pump phase $\Delta\phi^{-}$ [Equation (\ref{Case3})] on $K^{2}(z)$. Figures \ref{F3}a, \ref{F3}b and \ref{F3}c show the result for a homogeneous, parabolic and square-root coupling profile, respectively. Figure \ref{F3}c shows the result for a coupling strength three times lower than that used in Figures \ref{F3}a and \ref{F3}b. The squeezed eigenvalues are degenerate two by two for the $m$th and $(N+1-m)$th nonlinear supermodes. We refer to them as side nonlinear supermodes. Likewise, the zero nonlinear supermode \David{$[m=l\equiv (N+1)/2)]$} is the only nondegenerate supermode and it is always efficiently built-up and squeezed, independently of the value of $\Delta\phi^{-}$ (solid, green). The oscillatory and hyperbolic limit cases we pointed out for propagation in Equations (\ref{k-sol}) and (\ref{l-sol}) explain and match the squeezing behaviors displayed in each case. Full degeneracy and efficient squeezing --hyperbolic-- is obtained for all the supermodes for $\Delta\phi^{-}=\pi/2$ (solid, green). $\Delta\phi^{-}=0$ produces oscillatory squeezing (solid, blue and orange) in the side nonlinear supermodes which decreases as the coupling strength $C_{0}$ increases (Figures \ref{F3}a-\ref{F3}c). Notably, for intermediate cases $\Delta\phi^{-}=\pi/8$ (dotted), $\pi/4$ (dashed), $3\pi/8$ (dot-dashed), squeezing builds up smoothly for the side nonlinear supermodes and it approaches degeneracy for long propagation distances, whereas at short distances it is disturbed by the oscillatory part of Equation (\ref{Case3}). However, this disturbance is important since it mixes the individual downconverted modes and thus triggers quantum correlations and entanglement in the individual basis as we show in section \ref{Clusters}. 

\begin{figure}[t]
  \centering
    \subfigure{\includegraphics[width=0.445\textwidth]{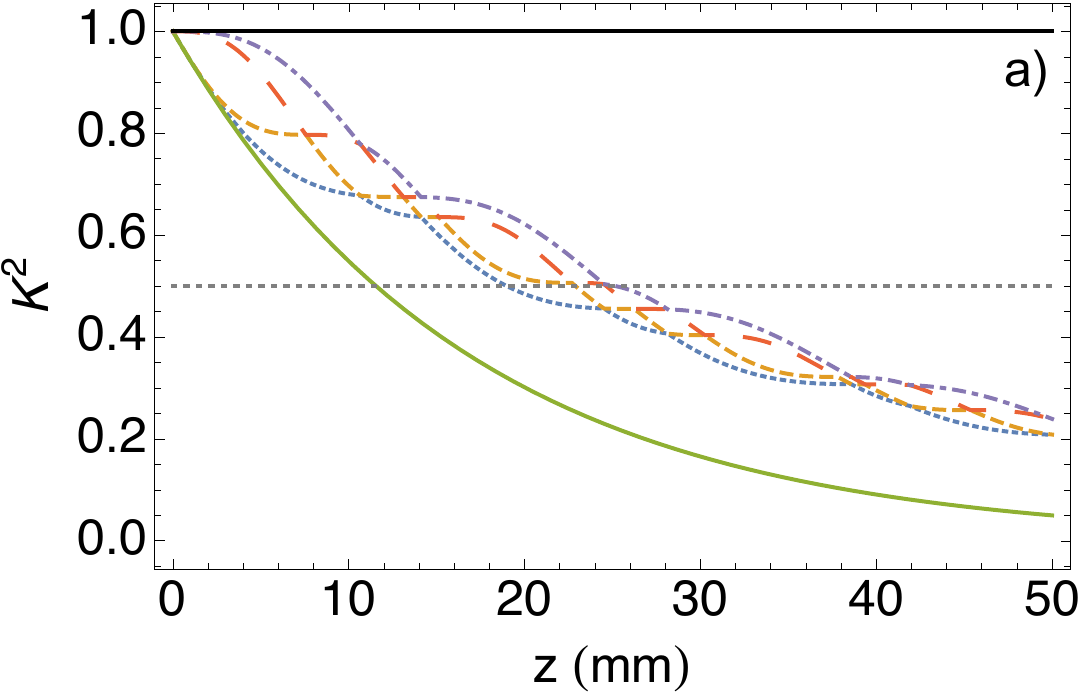}}
\vspace {0cm}\,
    \subfigure{\includegraphics[width=0.445\textwidth]{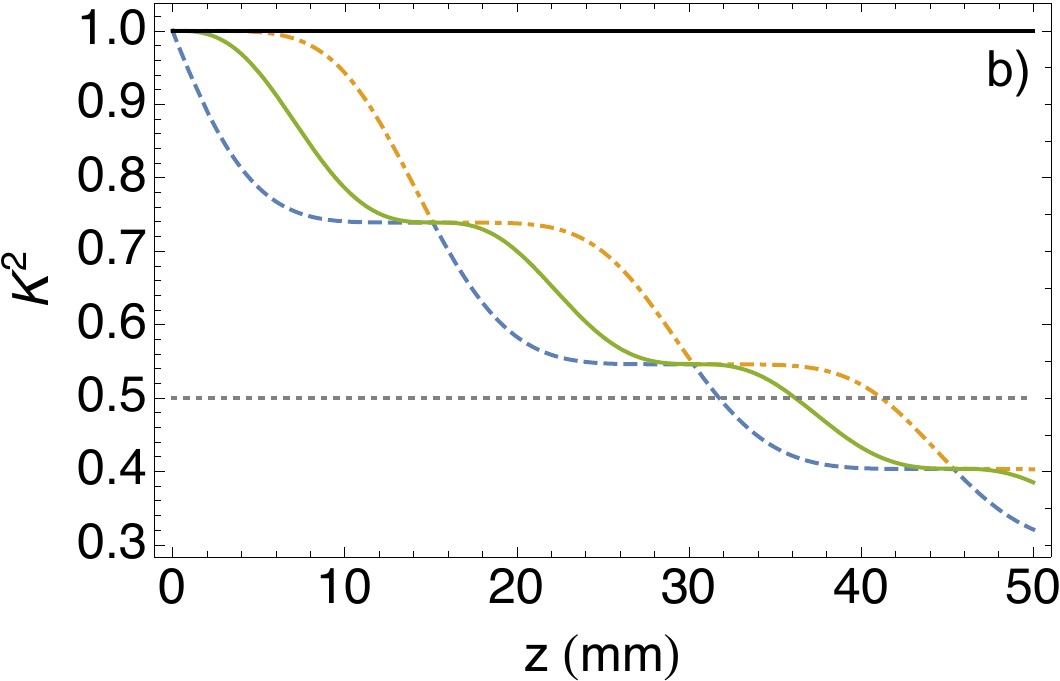}}
\vspace {0cm}\,
\hspace{0cm}\caption{\label{F4}\small{Evolution of nonlinear supermode squeezing $K^{2}_{N+m}(z)$ in a five-waveguides homogeneous coupling-profile nonlinear array pumping only a) the odd waveguides and b) the central waveguide. \David{a) The zero supermode $m=3$ is in solid green, and the side supermodes $m=1,2,4$ and $5$ are respectively in dotted blue, dashed orange, large-dashed red and dot-dashed violet. b) The zero supermode $m=3$ is in solid green and the side supermodes $m=1$ and $5$ are respectively in dashed blue and dot-dashed orange. The side supermodes $m=2$ and $4$ are in vacuum (black).} 3 dB squeezing level in dotted gray. $C_{0}=0.24$ mm$^{-1}$. $\eta=0.015$ mm$^{-1}$.}}
\end{figure}

{\itshape b. Influence of the coupling strength and coupling profile on multimode squeezing}. In addition to the effect of increased homogeneous coupling on oscillatory behavior exemplified by the comparison of Figures \ref{F3}a and \ref{F3}c, we display the different features obtained for homogeneous and parabolic coupling profiles (Fig. \ref{F3}a-\ref{F3}b). For $\Delta\phi^{-}=0$, there are certain lengths for the parabolic coupling profile where only the zero nonlinear supermode survives due to the equal spacing between the supermode propagation constants. Remarkably, for even number of waveguides and a parabolic coupling profile (not shown), there are propagation distances where destructive interference destroys all the SPDC generated light due to a evolving phase mismatch that periodically switches the system from downconversion to upconversion. Recently, bipartite entanglement between non-coupled pump fields has been demonstrated through this effect for two waveguides in the optical parametric amplification and second harmonic generation regimes \cite{Barral2017b, Barral2018}. Thus, this effect can also produce multipartite entanglement between non-interacting fields. We outline that the parabolic-coupling profile excited with a flat pump profile represents the spatial analogous case to the case of a frequency comb pumped with a Gaussian spectral shape since the Krawtchouk supermodes are Hermite-Gaussian functions in the continuous limit \cite{Patera2010}. 

{\itshape c. Linear vs nonlinear supermodes for a flat pump profile with uniform phase}. {We would like to end this section with a small discussion on the connection between solutions in terms of linear and nonlinear supermodes. To that end we use the flat pump configuration with uniform phase $\Delta\phi^{-}=0$.} The nonlinear supermodes diagonalize the covariance matrix as shown in Figures \ref{F2}a and \ref{F2}g. However, in that case the linear supermodes also diagonalize the covariance matrix as shown in section \ref{Engineerb1}. Both basis exhibit the same levels of squeezing, but different spatial profile evolution \cite{Note0}. The spatial profile related to the zero nonlinear supermode \David{$[m=l\equiv (N+1)/2)]$} obtained from $R_{1}(z)$ coincides with that calculated with Equation (\ref{l-sol}), but the side nonlinear supermodes \David{$(m\neq l)$} are slightly different from the side linear supermodes \David{$(k\neq l)$} obtained through Equation (\ref{k-sol}) and change with propagation. The cause of this disagreement is that the flat pump configuration diagonalizes the system up to a local phase rotation --the covariance matrix in the linear supermode basis is block-diagonal--, i.e. the quadratures of the linear supermodes are not at the maximum and minimum of the squeezing ellipse, whereas the Bloch-Messiah decomposition yields a fully diagonal covariance matrix. From the point of view of the experiment, this phase does not make any difference since the local oscillator of the balanced homodyne detector will sweep the entire squeezing ellipse. However, the linear supermode approach is here far more insightful and practical than Bloch-Messiah's one since the spatial profile is invariant along propagation and thus the $k$th supermode squeezing can be measured with a fixed LO profile $\vec{\theta}_{k}=\{M_{k, 1}, M_{k,2},\dots, M_{k,N}\}$, \David{whereas the LO profile used to measure the $m$th nonlinear supermode squeezing} would depend indeed on the length of the sample, the pump power and the coupling strength. More details on this are found in the Appendix \ref{apA}.

\subsubsection{Evolution of multimode squeezing with non-flat pumping profile}

{\itshape a. Pumping every other waveguide}. Figure \ref{F4}a shows the evolution of noise squeezing when pumping only the odd waveguides ($\vert \eta_{2j-1}\vert=\vert\eta\vert$, $\vert\eta_{2j}\vert=0$) of a $N=5$ waveguides homogeneous coupling-profile nonlinear array. This pump configuration indeed excites the five nonlinear supermodes, with a zero nonlinear supermode efficiently squeezed (green) and side nonlinear supermodes squeezing building up hyperbolically with an oscillatory modulation. The analytical solution obtained for the linear supermodes Equation (\ref{BEO}) anticipated this feature, since the side supermodes $(k, N+1-k)$ are coupled two by two in that basis. We outline that in the case of pumping the even waveguides ($\vert \eta_{2j}\vert=\vert\eta\vert$, $\vert\eta_{2j-1}\vert=0$), we obtain the same solution for the side nonlinear supermodes but with the zero nonlinear supermode in vacuum state. This is due to the inability to excite a supermode composed of odd elements when pumping the even waveguides in arrays with symmetric coupling profiles. Thus, in terms of multimode squeezing as a resource for quantum information the odd pumping is more efficient. Note that this does not happen in arrays with asymmetric coupling profiles like the square-root coupling profile (see Figure \ref{F1a}c). In the case of an ANW made up of an even number of waveguides we excite all the supermodes independently on the parity of the total number of waveguides since there is no zero supermode.

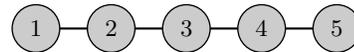
\begin{figure}[t]
\begin{center}
\begin{tikzpicture}
  \GraphInit[vstyle=Dijkstra]
  \SetVertexNormal[Shape=circle,FillColor=black!20]
  \Vertex[x=0,y=0.707,L=$1$]{A}    
  \Vertex[x=1,y=0.707,L=$2$]{B}    
  \Vertex[x=2,y=0.707,L=$3$]{C}
  \Vertex[x=3,y=0.707,L=$4$]{D}
  \Vertex[x=4,y=0.707,L=$5$]{E}
  \tikzset{EdgeStyle/.style={-}}
  \Edge (A)(B)
  \Edge (B)(C)
  \Edge (C)(D)
  \Edge (D)(E)
\end{tikzpicture}
\end{center}
\caption{\label{F5}\small{Graph corresponding to 5-mode linear cluster state.}}
\end{figure}

{\itshape b. Pumping the central waveguide in an odd ANW}. Figure \ref{F4}b shows the evolution of noise squeezing when pumping only the central waveguide ($\vert \eta_{j}\vert= \vert \eta \vert\, \delta_{j,l}$) of a $N=5$ waveguides homogeneous coupling-profile nonlinear array. This pump configuration leads to the excitation of only three out of five nonlinear supermodes --the odd ones--, with the other two in vacuum state along propagation. The comparison between Figure \ref{F3}a and Figure \ref{F4}b thus further sheds light on the difference between Figure \ref{F2}g and Figure \ref{F2}i. The squeezing increases hyperbolically with an oscillatory modulation. Note that in this pumping configuration there is no direct correlation between linear and nonlinear supermodes anymore. However, the nonlinear supermode squeezing exhibited in this case can be also explained in terms of the linear supermodes as in Section \ref{Engineer}.B.5. The leading terms of the zero \David{($k=l=3$)} and side \David{($k=1,5$)} linear supermode equations correspond to degenerate and nondegenerate parametric amplifiers, respectively, leading to hyperbolic squeezing. First-order terms introduce a z-dependent coupling between the zero and the side supermodes with period $z_{p}\approx 2\pi/\vert \lambda_{1 (5)}\vert$ for $\vert \lambda_{1 (5)}\vert \gg 2 \vert \eta\vert$. The main difference here with respect to the flat pump profile case is that with central pumping the linear supermodes do not evolve independently but together, leading to coupling. The Bloch-Messiah decomposition evidences the squeezing arising from this effect, as exhibited in Figure \ref{F4}b. The period shown in the figure agrees with that calculated $z_{p}=2\pi/(\sqrt{3} \,C_{0})=15.1$ mm. The level of squeezing is lower than that obtained in Figure \ref{F3} at the same distance since we use the same input pump power per waveguide, but the total power available per individual mode is $1/5$. We finally outline that this configuration, being the more simple in terms pumping profile, is not an efficient resource for quantum information since only a subset of the available nonlinear supermodes is squeezed.

In conclusion of this section \ref{Multimode}, we have demonstrated how the insight gained in section \ref{Engineer} from the mode propagation can be used to engineer multipartite squeezing in ANW. The tuning of the pumping and coupling parameters, together with suitable encoding leads to different configurations of squeezing. The multimode squeezing presented above is a resource for multimode entanglement. We have indeed demonstrated very recently a protocol for the generation of large multimode entangled states for quantum networks \cite{Barral2020b} and the versatile production of cluster states for quantum computing in ANW \cite{Barral2020}. In the next section we focus on a specific class of entangled states useful for quantum computing --the cluster states \cite{Raussendorf2001}. We show that pumping with a flat profile as introduced in section \ref{Engineer} is a good strategy to generate large linear cluster states. In particular we demonstrate how to choose a good working point in an analytically and semi-analytically scanned parameter space and how to further numerically optimize the parameters.

\section{Efficient generation of linear cluster states} \label{Clusters}

An ideal CV cluster state is a simultaneous eigenstate of specific quadrature combinations called nullifiers \cite{Raussendorf2001, Menicucci2006}. Cluster states are associated with a graph or \Da{adjacency matrix $J$}. The nodes of the graph represent the modes of the cluster state in a given basis, and the edges the entanglement connections among the nodes. Moreover, the label of the modes that are part of the cluster can be suitably set to maximize the entanglement between nodes. The nullifiers are given by
\begin{equation}\label{Nul}\nonumber
\hat{\delta}_{i}\equiv \hat{x}_{i}(\theta_{i}+\pi/2)-\sum_{l=1}^{N} J_{i, i'}\, \hat{x}_{i'}(\theta_{i'}) \quad \forall i=1,\dots, N,
\end{equation}
where $J$ is the graph associated to the cluster and $\hat{x}_{i}({\theta_{i}})=\hat{x}_{i} \cos{(\theta_{i})}+\hat{y}_{i} \sin{(\theta_{i})}$ is the $i$th generalized quadrature in a given basis. We consider unit-weight cluster states with $J_{i,i'} = 1$ for modes $i$ and $i'$ being nearest neighbors in the graph and all the other entries of $J$ are zero. 

Cluster states are the resource of CV measurement-based quantum computing (MBQC) \cite{Gu2009}. The computation relies in this framework on the availability of a large multimode entangled state on which a specific sequence of measurements is performed. The choice of basis widens the range of application in MBQC \cite{Ferrini2013}. The nullifier variances tend to zero in the ideal limit of infinite squeezing. Experimentally, a cluster state can be certified if two conditions are satisfied: i) the noise of a a set of normalized nullifiers lies below shot noise
\begin{equation}\label{VNul}\nonumber
V(\bar{\delta}_{i})<1 \quad \forall i=1,\dots, N,
\end{equation}
where $\bar{\delta}_{i}\equiv \delta_{i}/\sqrt{1+n(i)}$ is the normalized nullifier and $n(i)$ is the number of nearest neighbors to the $i$th node of the cluster, and ii) the cluster state is fully inseparable, i.e. it violates a set of multipartite entanglement inequalities \cite{vanLoock2003, Yukawa2008}.

We exhibit here how linear cluster states encoded in the individual mode basis are produced naturally in the flat pump configuration -- introduced in section \ref{Engineer} and explored from a squeezing point of view in section \ref{Multimode}. Notably, in the context of MBQC, a linear 4-mode cluster state is a sufficient resource for an arbitrary single-mode Gaussian unitary \cite{Ukai2010}. Hence, linear cluster states represent key resources in this domain. The adjacency matrix $J_{lin}$ corresponding to a linear cluster is the same as that related to the coupling in a homogeneous array when the encoding $i$th node = $j$th mode is used. Thus, the ANWs can be a natural platform for the generation of this class of cluster states.

\begin{figure}
\centering
\subfigure{\includegraphics[width=0.45\textwidth]{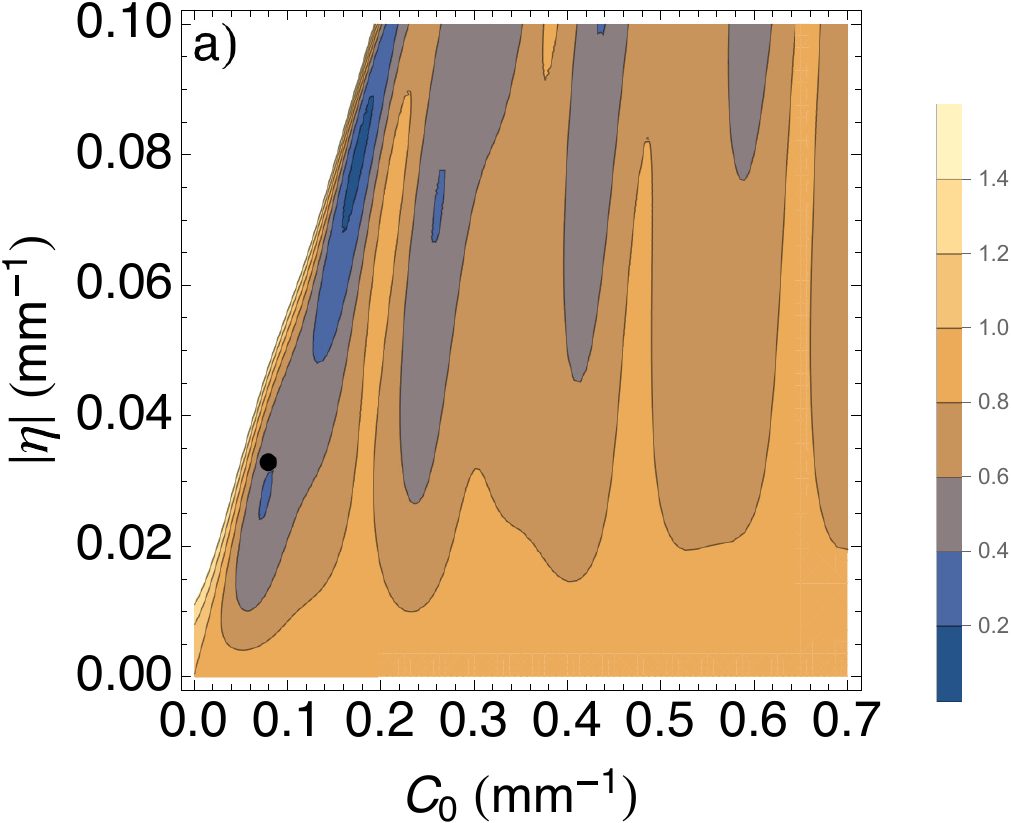}} 
    \quad \subfigure{\includegraphics[width=0.45\textwidth]{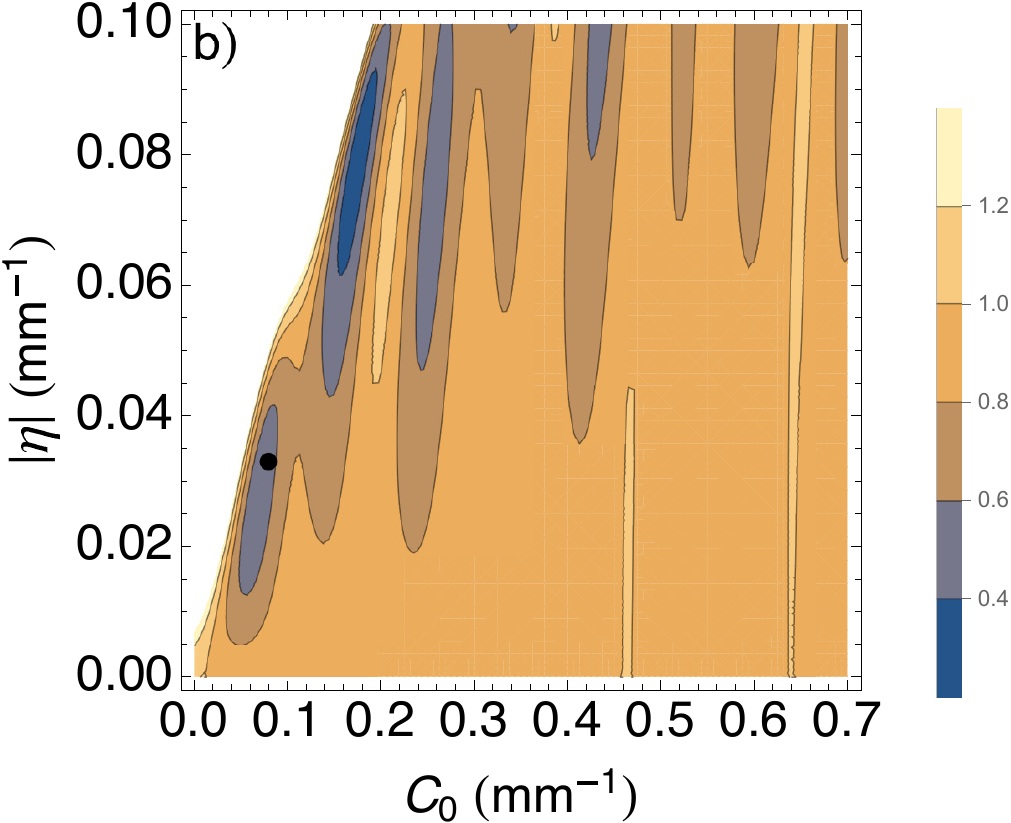}}
     \quad \subfigure{\includegraphics[width=0.45\textwidth]{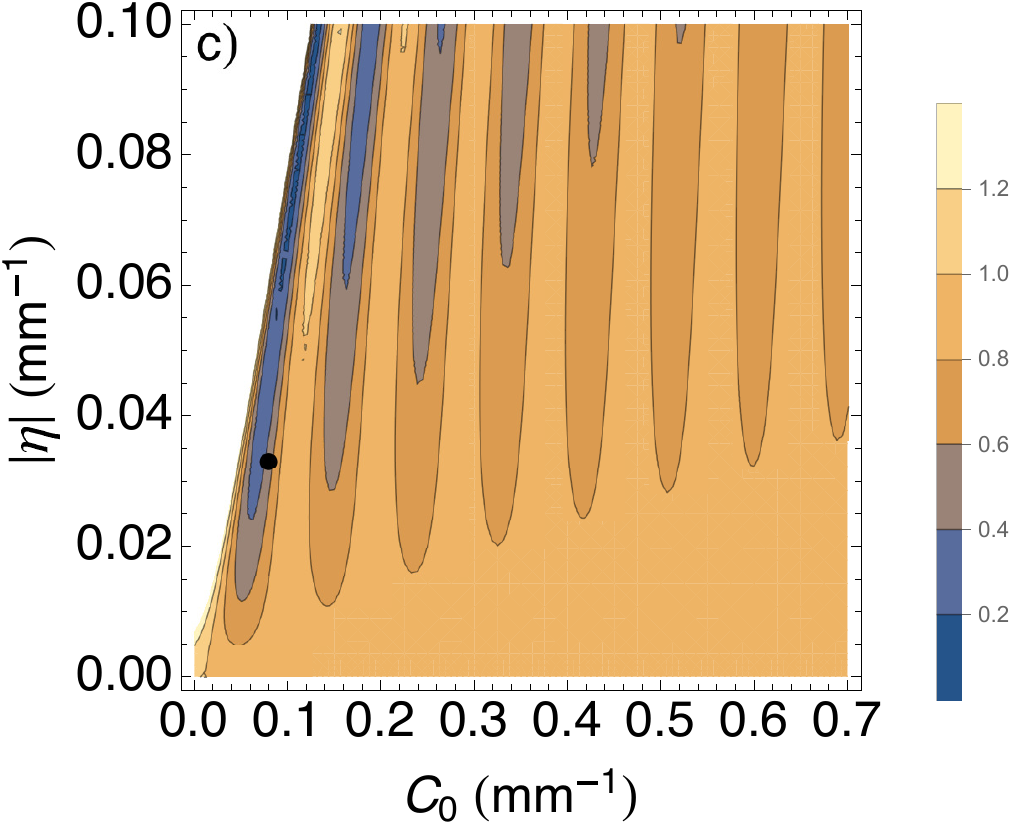}}
\vspace {0cm}
\hspace{0cm}\caption{\label{F8a}\small{\David{Nullifier variances} $V(\bar{\delta}_{i})$ for a $N=5$ linear cluster state generated in an ANWs with homogeneous coupling and flat pump profiles \David{as a function of the coupling strength $C_{0}$ and the pump power via $\vert \eta \vert$}. Simultaneous values of $V(\bar{\delta}_{i}) < 2/3$ are signature of  cluster production. a) $V(\bar{\delta}_{1})=V(\bar{\delta}_{5})$, b) $V(\bar{\delta}_{2})=V(\bar{\delta}_{4})$, and c) $V(\bar{\delta}_{3})$. The white areas stand for $V(\bar{\delta}_{i}) \gg 1$. The black dot marks the point $(C,\eta)=(0.08, 0.033)$ mm$^{-1}$. $\phi_{j}=-\pi/2$. $z=20$ mm.   }}
\end{figure}

The normalized nullifiers for a linear cluster composed of $N$ modes are given by
\begin{equation}\nonumber
\bar{\delta}_{i}=\frac{y_{i} (\theta_{i}) - x_{i-1}(\theta_{i-1}) - x_{i+1}(\theta_{i+1})}{\sqrt{1+n(i)}},
\end{equation}
with $x_{0}(\theta_{0})=x_{N+1}(\theta_{N+1})=0$ and where we have related the $i$th node of the cluster with the $j$th individual mode of the ANWs ($i=j=1, \dots, N$). The full inseparability of the cluster nodes can also be assessed by means of the Van Loock - Furusawa multipartite entanglement witness (VLF) \cite{vanLoock2003}. For a linear cluster the VLF is given in terms of the following $N-1$ inequalities \cite{Yukawa2008}
\begin{equation}\label{VLFlinear}\nonumber
V(\bar{\delta}_{i})+V(\bar{\delta}_{i+1})\geq
\begin{cases}   
\sqrt{\frac{8}{3}} \quad &\text{for $i=1, N-1$},\\ 
\quad \frac{4}{3} \quad \quad &\text{for $i=2, \dots, N-2$}.
\end{cases}
\end{equation}
Thus, simultaneous values of $V(\bar{\delta}_{i}) < 2/3$ ensure the production of a linear cluster. Below we exhibit the use of the analytical solutions Equations (\ref{V0}) in the generation of linear cluster states as that shown in Figure \ref{F5}. These solutions are a suitable initial working point in the huge space of parameters for the production of a linear cluster. This is very important in order to design a sample: we have a starting point from which optimization via pumping profile, LO phases and electronic gains can improve the result. Remarkably, we have found that these solutions are not very far from the best working point to produce this kind of states \cite{Barral2020}.

Figure \ref{F8a} maps the nullifier variances characterizing a $N=5$ linear cluster state produced in an ANWs with homogeneous coupling and propagation length $z=20$ mm. Due to the symmetry of the system the nullifiers are degenerate two by two, but the $l$th nullifier:  $V(\bar{\delta}_{1})=V(\bar{\delta}_{5})$ (Figure \ref{F8a}a), $V(\bar{\delta}_{2})=V(\bar{\delta}_{4})$ (Figure \ref{F8a}b), and $V(\bar{\delta}_{3})$ (Figure \ref{F8a}c). The contour plots display common areas fulfilling the condition $V(\bar{\delta}_{i}) < 2/3$ (blue areas). For instance, for $\vert \eta \vert=0.06$ mm$^{-1}$ and $C_{0}=0.16$ mm$^{-1}$, we get  $V(\bar{\delta}_{1(5)})=0.34$, $V(\bar{\delta}_{2(4)})=0.42$, and $V(\bar{\delta}_{3})=0.40$. These values are of the order of those obtained in the frequency domain with frequency combs \cite{Medeiros2014}.

\begin{figure}[t]
  \centering
    \subfigure{\includegraphics[width=0.45\textwidth]{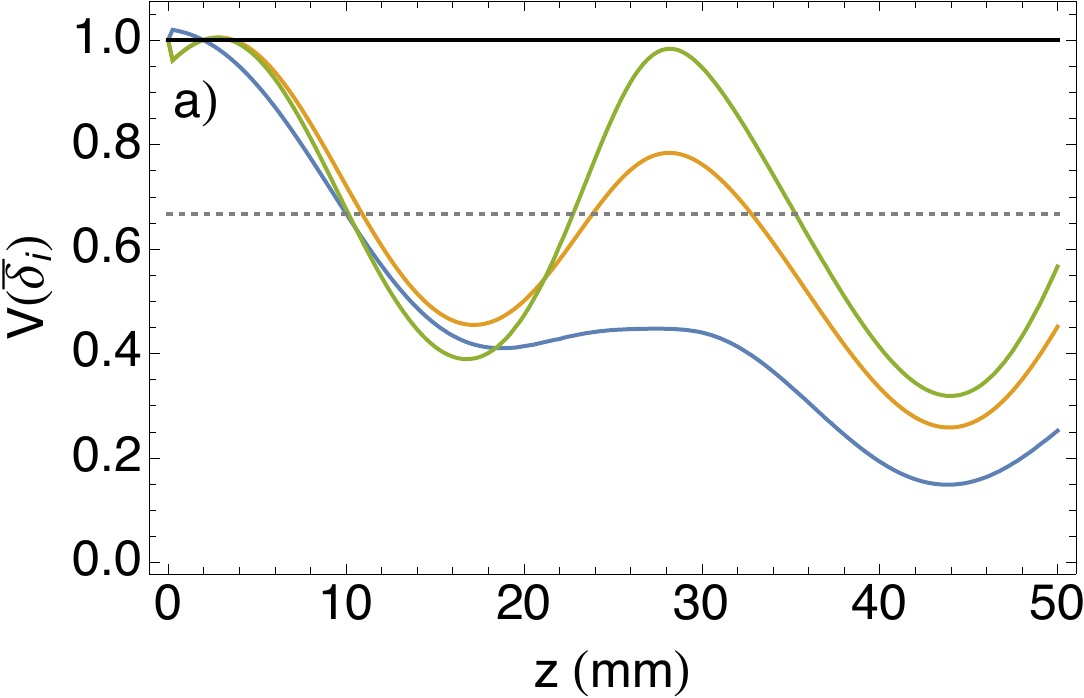}}
          \vspace{0cm}
      \subfigure{\includegraphics[width=0.45\textwidth]{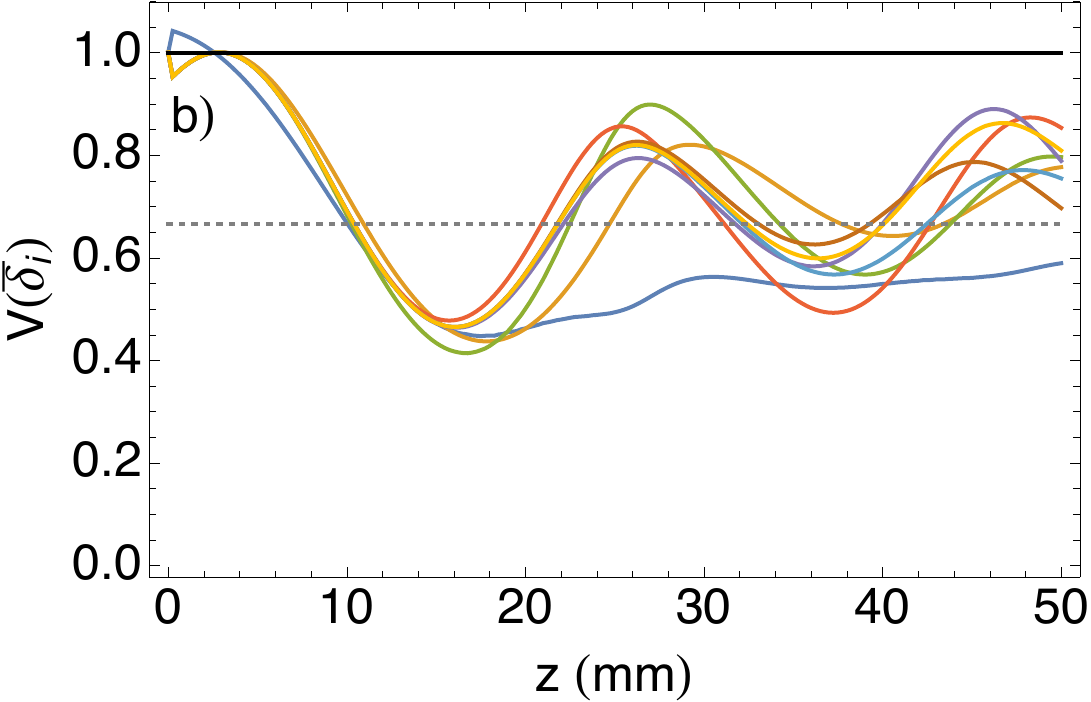}}  
  \vspace {0cm}\,
\hspace{0cm}\caption{\label{F8b}\small{\David{Nullifier variances $V(\bar{\delta}_{i})$ for a) $N=5$ and b) $N=15$ linear cluster states generated in an ANWs with homogeneous coupling and flat pump profiles as a function of the propagation length $z$}. 
Simultaneous values of $V(\bar{\delta}_{i}) < 2/3$ (dotted, gray) are signature of cluster production. (a) $N=5$ with, from lower to upper at $z=50$ mm, $i=1,5$ (blue), $i=2,4$ (orange), and  $i=3$ (green). (b) $N=15$ with, from lower to upper at $z=50$ mm, $i=1,9$ (blue), $i=6, 14$ (brown), $i=7, 15$ (sky blue), $i=2,10$ (orange), $i=5, 13$ (purple), $i=3, 11$ (green), $i=3$ (yellow), and $i=4, 12$ (red). $C_{0}=0.08$ mm$^{-1}$. $\phi_{j}=-\pi/2$. }}
\end{figure}

In order to gain insight about the scalability of this configuration, Figure \ref{F8b} pictures the evolution along propagation of the nullifier variances related to linear cluster states made up of $N=5$ (Figure \ref{F8b}a) and $N=15$ (Figure \ref{F8b}b) modes. Now, we optimize the amount of power per waveguide $\eta$ for a given coupling constant. We use the sum of the five (fifteen) nullifier variances $F_{C}(\eta)=\sum_{i=1}^{5(15)} V(\bar{\delta}_{i})$ at each $z$ as the fitness function to optimize. We use an evolution-strategy algorithm to tackle this optimization \cite{Beyer2002}. As commented above, the nullifier variances are degenerate due to the symmetry of the system. Remarkably, the linear cluster condition $V(\bar{\delta}_{i}) < 2/3$ is fulfilled in both cases for a large range of distances. In order to connect Figures \ref{F8a} and \ref{F8b}, we have marked as a black dot in Figures \ref{F8a}a, b and c, the coordinates $(C, \eta)=(0.08, 0.033)$ mm$^{-1}$ corresponding to the variances of the nullifiers at $z=20$ mm shown in Figure \ref{F8b}a. The maxima values of $\eta$ used in the optimization are 0.038 and 0.035 mm$^{-1}$ for $N=5$ and 15, respectively. These values are attainable with current technology \cite{Mondain2019, Takanashi2020, Kashiwazaki2020}. Note that the coupling constant is wavelength dependent $C_{0}=C_{0}(\omega_{s})$ \cite{Kruse2015}. Thus, for a fixed ANW length, modifying the operating wavelength $\lambda_{s}$ and the temperature of the sample $T$, we can access to more favorable conditions to obtain multipartite entanglement. This is clearly shown in Figure \ref{F8a} when fixing the value of nonlinear strength $\vert\eta\vert$ and checking the value of the nullifiers for different values of coupling strength $C_{0}$.

We have demonstrated the production of linear cluster states with our analytical solutions Equations (\ref{V0}). However, the parameter space of the full approach is much larger than that corresponding to this special case. This enables the optimized generation of linear and other classes of cluster states in the individual mode basis or any other basis \cite{Barral2020}.

\section{conclusions}

\David{\mw{ The ANW is a versatile system for quantum state engineering, as presented Barral et al., Phys. Rev. Appl. $\bf{14}$, 044025 (2020). Here, we complemented this claim, based on optimization of parameters for the generation of specific multimode entangled states, with a more fundamental perspective through a comprehensive analysis of scalable analytical solutions. As such, we build an intuition for the possibilities of the system in terms of the available tuning parameters. This analytical approach provides insight into the particular features of the multimode squeezed states, produced in a given parameter configuration, and into working points in the parameter-space which maximize given properties of the generated state.}

\mw{In particular, we have unfolded how the available internal and external parameters of the ANW affect the generated multimode squeezing, which is at the root of multimode entanglement.} We have especially detailed how the fields propagate and how their fluctuations are squeezed in three relevant \mw{mode} bases, related to three encodings of quantum information. The practical individual \mw{mode} basis, where each mode corresponds to an individual waveguide, naturally provides individual outputs \mw{which are} useful in quantum networks architectures\mw{. The} linear supermode basis provides insight through analytical solutions and simple detection with a constant LO profile\mw{. Finally,} the local nonlinear supermode basis maximizes the squeezing resources and is instrumental for multimode entanglement through LO shaping or postprocessing. We have provided insights on the engineering choices that can be made in terms of coupling strength and profile, measurement
strategy and pumping geometry in phase and amplitude. We have exemplified the usefulness of our analytical solutions and further numerical optimization providing working points in the parameter space to produce linear cluster and assess their scalability. The quantum information-encoding strategies and extended toolbox, \mw{that are} provided here, are applicable to all implementation\mw{s} of nonlinear waveguide arrays including very recent and promising development\mw{s} \cite{Mondain2019, Takanashi2020, Kashiwazaki2020, Boes2018, Loncar2018}. The analyzed compact and original interplay of nonlinearity and coupling in the nonlinear arrays of waveguides produces multimode entanglement in a way that is accessible only to guided-wave nonlinear components and serve the purpose of quantum technologies. \mw{Our results thus provide an innovation to implement quantum protocols with integrated optics in a compact way.}}

\section*{Acknowledgements} The authors thank G. Patera for useful comments. This work was supported by the Agence Nationale de la Recherche through the INQCA project (Grants No. PN-II-ID-JRP-RO-FR-2014-0013 and No. ANR-14-CE26-0038), the Paris Ile-de-France region in the framework of DIM SIRTEQ through the project ENCORE, and the Investissements d'Avenir program (Labex NanoSaclay, reference ANR-10-LABX-0035).

\appendix*

\section{}\label{apA}
The covariance matrix in the linear supermode basis $V_{S}$ obtained directly from Equation (\ref{k-sol}) is block diagonal. Suitable rotations (phase shifts) in the phase space related to each supermode can however diagonalize fully the linear supermode covariance matrix $V_{S}$. From Equation (\ref{k-sol}), we can straightforwardly calculate the covariance matrix $V_{S}$ related to the uncoupled $k$th linear supermode. A rotation in the $k$th supermode phase space of an angle 
\begin{equation}\nonumber
\vartheta_{k}=\frac{1}{2} \arctan{[\frac{2V(x_{S,k}, y_{S,k})}{V(y_{S,k}, y_{S,k})-V(x_{S,k}, x_{S,k})}]} + \frac{\pi}{2},
\end{equation}
diagonalizes the covariance matrix $V_{S}$ yielding
\begin{align} \nonumber
&V(x_{S,k}', x_{S,k}')=  \frac{V(x_{S,k}, x_{S,k})+V(y_{S,k}, y_{S,k})}{2} \\   \nonumber
&\qquad+\frac{\sqrt{(V(y_{S,k}, y_{S,k})-V(x_{S,k}, x_{S,k}))^{2}+4 V(x_{S,k}, y_{S,k})^{2}}}{2},\\      \nonumber
&V(y_{S,k}', y_{S,k}')=  \frac{V(x_{S,k}, x_{S,k})+V(y_{S,k}, y_{S,k})}{2} \\ \nonumber
&\qquad-\frac{\sqrt{(V(y_{S,k}, y_{S,k})-V(x_{S,k}, x_{S,k}))^{2}+4 V(x_{S,k}, y_{S,k})^{2}}}{2}.
\end{align}
This diagonal matrix is the same as $K^{2}(z)$ obtained by the Bloch-Messiah decomposition Equation (\ref{BMD}). $R_{1}(z)$ can be factorized thus as $R_{1}(z)=M R(\vec{\vartheta})$, with
\begin{equation}\label{Rtheta} \nonumber
 \quad R(\vec{\vartheta})=\begin{pmatrix} \cos{(\vec{\vartheta})} & \sin{(\vec{\vartheta})} \\ -\sin{(\vec{\vartheta})} & \cos{(\vec{\vartheta})} \end{pmatrix},
\end{equation}
and $\cos{(\vec{\vartheta})}=\diag\{\cos{(\vartheta_{1})}, \dots, \cos{(\vartheta_{k})}, \dots, \cos{(\vartheta_{N})}\}$ [equally for $\sin{(\vec{\vartheta})}$]. 

We show an example for the sake of clarification. For a pump phase profile $\Delta\phi^{-}=0$ with $\phi=0$, the covariance matrix elements in the propagation supermode basis are
\begin{align} \nonumber
V(x_{S,k}, x_{S,k})&= [\cosh{(r_{k})} + \sinh{(r_{k})} \cos{(2 F_{k} z)}] e^{-r_{k}},\\     \nonumber
V(y_{S,k}, y_{S,k})&= [\cosh{(r_{k})} - \sinh{(r_{k})} \cos{(2 F_{k} z)}] e^{r_{k}}, \\      \nonumber
V(x_{S,k}, y_{S,k})&= \sinh{(r_{k})} \sin{(2 F_{k} z)},
\end{align}
with $r_{k}=(1/2) \ln{[(\lambda_{k}+2\vert \eta \vert)/(\lambda_{k}-2\vert \eta \vert)]}$. 
The squeezing phase is given by $\vartheta_{k}=\pi/2 - (1/2) \arctan{\{[(\cosh{(r_{k})} \tan{(F_{z} z)}]^{-1}\}}$. It depends on the pump power via $\vert \eta \vert$, the coupling strength $C_{0}$ and the propagation length $z$ --and thus $R_{1}(z)$--. The larger squeezing is obtained periodically at distances $z_{k}=(2n+1)\pi/(2 F_{k})$ different for each $k$th supermode, with $n$ any positive integer. The diagonalized variances at those distances are $V(x_{S,k\neq l}', x_{S,k \neq l}')=e^{2 r_{k}}$, $V(y_{S,k \neq l}', y_{S,k \neq l}')=e^{-2 r_{k}}$. These are the same values as the minima of the blue and orange curves in Figure \ref{F3}. The $k$th mode squeezing disappears at periodic distances $z_{k}'=n\pi/F_{k}$, the maxima of blue and orange curves in Figure \ref{F3}. In the case of an odd number of waveguides, a $\vartheta_{l}=\pi/4$ rotation in phase space diagonalizes the covariance matrix corresponding to the zero supermode independently of z, with $V(x_{S,l}', x_{S,l}')=e^{4\vert \eta \vert z}$, $V(y_{S,l}', y_{S,l}')=e^{-4\vert \eta \vert z}$ (green curves in Figure \ref{F3}). The zero supermode is therefore the same for both bases. 

In summary, the total available squeezing of the linear and nonlinear supermodes is the same, but it is distributed in a different way.


\section*{Bibliography}
\begin{thebibliography}{80}

\bibitem{Braunstein2005} S.L. Braunstein and P. van Loock. {\it Rev. Mod. Phys.} {\bf 77}, 513 (2005).
\bibitem{Ou1992} Z.Y. Ou, S.F. Pereira, H.J. Kimble and K.C. Peng. {\it Phys. Rev. Lett.} {\bf 68}(25), 3663-3666 (1992).
\bibitem{Furusawa1998} A. Furusawa, J.L. Sorensen, S.L. Braunstein, C.A. Fuchs, H.J. Kimble and E.S. Polzik. {\it Science} {\bf 282}, 706 (1998).
\bibitem{Jia2004} X. Jia, X. Su, Q. Pan, J. Gao, C. Xie and K. Peng. {\it Phys. Rev. Lett.} {\bf 93}, 250503 (2004).
\bibitem{Coelho2009} A.S. Coelho, F.A.S. Barbosa, K.N. Cassemiro, A.S. Villar, M. Martinelli and P. Nussenzveig. {\it Science} {\bf 326}, 823 - 826 (2009).
\bibitem{Miwa2009} Y. Miwa, J. Yoshikawa, P. van Loock and A. Furusawa. {\it Phys. Rev. A} {\bf 80}, 050303(R) (2009).
\bibitem{Jouguet2013} P. Jouguet, S. Kunz-Jacques, A. Leverrier, P. Grangier and E. Diamanti. {\it Nature Photon.} {\bf 7}, 378 (2013).
\bibitem{Larsen2019} M.V. Larsen, X. Guo, C.R. Breum, J.S. Neergaard-Nielsen and U.L. Andersen. {\it Science} {\bf 366}, 369 - 372 (2019).
\bibitem{Asavanant2019} W. Asavanant, Y. Shiozawa, S. Yokoyama, B. Charoensombutamon, H. Emura, R.N. Alexander, S. Takeda, J. Yoshikawa, N. C. Menicucci, H. Yonezawa and A. Furusawa. {\it Science} {\bf 366}, 373 - 376 (2019).
\bibitem{Wang2019} J. Wang, F. Sciarrino, A. Laing and M.G. Thompson. {\it Nature Phot.} {\bf 14}, 273-284 (2020).

\bibitem{Lenzini2018} F. Lenzini, J. Janousek, O. Thearle, M. Villa, B. Haylock, S. Kasture, L. Cui, H.-P. Phan, D.V. Dao, H. Yonezawa, P.K. Lam, E.H. Huntington and M. Lobino. {\it Science Advances} {\bf 4} (12), eaat9331 (2018).
\bibitem{Yukawa2008} M. Yukawa, R. Ukai, P. van Loock and A. Furusawa. {\it Phys. Rev. A} {\bf 78}, 012301 (2008). 

\bibitem{Christodoulides2003} D.N. Christodoulides, F. Lederer and Y. Silberberg. {\it Nature} {\bf424} (6950), 817-23 (2003).
\bibitem{Kruse2013} R. Kruse, F. Katzschmann, A. Christ, A. Schreiber, S. Wilhelm, K. Laiho, A. G\'abris, C.S. Hamilton, I. Jex and Ch. Silberhorn. {\it New. J. Phys.} {\bf 15}, 083046 (2013).
\bibitem{Solntsev2014} A.S. Solntsev, F. Setzpfandt, A.S. Clark, C.W. Wu, M.J. Collins, C. Xiong, A. Schreiber, F. Katzschmann, F. Eilenberger, R. Schieck, W. Sohler, A. Mitchell, Ch. Silberhorn, B.J. Eggleton, T. Pertsch, A.A. Sukhourukov, D.N. Neshev and Y.S. Kivshar. {\it Phys. Rev. X} {\bf 4}, 031007 (2014).
\bibitem{Herec2003} J. Herec, J. Fiurasek and L. Mista Jr. {\it J. Opt. B: Quantum Semiclass. Opt.} {\bf 5}, 419-426 (2003).
\bibitem{Barral2017b} D. Barral, N. Belabas, L.M. Procopio, V. D'Auria, S. Tanzilli and J.A. Levenson. {\it Phys. Rev. A} {\bf 96}, 053822 (2017).
\bibitem{Rai2012} A. Rai and D.G. Angelakis. {\it Phys. Rev. A} {\bf 85}, 052330 (2012). 
\bibitem{Barral2019b} D. Barral, K. Bencheikh, N. Belabas and J.A. Levenson. {\it Phys. Rev. A} {\bf 99}, 051801 (R) (2019).

\bibitem{Barral2020b} D. Barral, K. Bencheikh, J.A. Levenson and N. Belabas. {\it arXiv: 2005.07241}, (2020).
\bibitem{Barral2020} D. Barral, M. Walschaers, K. Bencheikh, V. Parigi, J.A. Levenson, N. Treps and N. Belabas. {\it Phys. Rev. Appl.} $\bf{14}$, 044025 (2020).

\bibitem{Roslund2013} J. Roslund, R. Medeiros de Araujo, S. Jiang, C. Fabre and N. Treps. {\it Nature Phot.} {\bf 8}, 109 - 112 (2013).
\bibitem{Cai2017} Y. Cai, J. Roslund, G. Ferrini, F. Arzani, X. Xu, C. Fabre and N. Treps. {\it Nature Comm.} {\bf 8}, 15645 (2017).
\bibitem{Arzani2018} F. Arzani, C. Fabre and N. Treps. {\it Phys. Rev. A} {\bf 97}, 033808 (2018).
\Da{
\bibitem{BenAryeh1991} Y. Ben-Aryeh and S. Serulnik. {\it Phys. Lett. A} {\bf 155}, 473 (1991).
\bibitem{Toren1994} M. Toren and Y. Ben-Aryeh. {\it Quantum Opt.} {\bf 6}, 425-444 (1994). 
\bibitem{Perina2000} J. Perina and J. Perina Jr. in {\it Progress in Optics} {\bf 41}, 361-419 (2000).
}


\bibitem{Chen2014} M. Chen, N.C. Menicucci and O. Pfister. {\it Phys. Rev. Lett.} {\bf 112}, 120505 (2014).
\bibitem{Su2012} X. Su, Y. Zhao, S. Hao, X. Jia, C. Xie, and K. Peng. {\it Opt. Lett.} {\bf 37}, 5178-5180 (2012). 
\bibitem{Armstrong2012} S. Armstrong, J.-F. Morizur, J. Janousek, B. Hage, N. Treps, P.K. Lam and H.-A. Bachor. {\it Nature Comm.}, 3:1026 (2012).
\bibitem{Armstrong2015} S. Armstrong, M. Wang, R.Y. Teh, Q. Gong, Q. He, J. Janousek, H.-A. Bachor, M.D. Reid and P.K. Lam. {\it Nature Phys.} {\bf 11}, 167-172 (2015).
\bibitem{Yoshikawa2016} J.-I. Yoshikawa, S. Yokoyama, T. Kaji, Ch. Sornphiphatphong, Y. Shiozawa, K. Makino and A. Furusawa. {\it APL Photonics} {\bf 1}, 060801 (2016).
\bibitem{Takeda2019} S. Takeda, K. Takase and A. Furusawa. {\it Science Advances} {\bf 5} (5), eaaw4530 (2019).
\bibitem{Patera2010} G. Patera, N. Treps, C. Fabre and G.J. Valcarcel. {\it Eur. Phys. J. D} {\bf 56}, 123-140 (2010).
\bibitem{Fabre2019} C. Fabre and N. Treps. {\it Rev. Mod. Phys.} {\bf 92}, 035005 (2020).

\bibitem{Kapon1984} E. Kapon, J. Katz and A. Yariv. {\it Opt. Lett.} {\bf 10} (4), 125-127 (1984).
\bibitem{Bosse2017} E.-O. Boss\'e and L. Vinet. {\it SIGMA} {\bf 13}, 074 (2017).
\Da{\bibitem{Christ2013} A. Christ, B. Brecht, W. Mauerer and Ch. Silberhorn. {\it New J. Phys.} {\bf 15}, 043038 (2013).}
\Da{\bibitem{Fiurasek2000} J. Fiurasek, J. Perina. {\it Phys. Rev. A} {\bf 62}, 033808 (2000).}
\Da{\bibitem{Quesada2014} N. Quesada and J.E. Sipe. {\it Phys. Rev. A} {\bf 90}, 063840 (2014).}

\bibitem{Cariolaro2016} G. Cariolaro and G. Pierobon. {\it Phys. Rev. A} {\bf 94}, 062109 (2016).
\bibitem{Hamilton2014} C.S. Hamilton, R. Kruse, L. Sansoni, Ch. Silberhorn and I. Jex. {\it Phys. Rev. Lett.} {\bf 113}, 083602 (2014).
\bibitem{Braunstein2005a} S.L. Braunstein. {\it Phys. Rev. A} {\bf 71}, 055801 (2005).
\bibitem{Adesso2014} G. Adesso, S. Ragy and A.R. Lee. {\it Open Syst. Inf. Dyn.} {\bf 21}, 1440001 (2014).
\bibitem{Simon1994} R. Simon, N. Mukunda and B. Dutta. {\it Phys. Rev. A} {\bf 49} (3), 1567-1583 (1994).
\bibitem{Barral2019c} D. Barral, N. Belabas, K. Bencheikh and J.A. Levenson. {\it Phys. Rev. A} {\bf 100}, 013824 (2019).
\bibitem{Vainio2011} M. Vainio and L. Halonen. {\it Opt. Lett.} {\bf 36} (4), 475-477 (2011).

\bibitem{Moison2009} J.M. Moison, N. Belabas, C. Minot and J.A. Levenson. {\it Opt. Lett.} {\bf 34}(16), 2462-2464 (2009).
\bibitem{Keil2011} R. Keil, A. Perez-Leija, F. Dreisow, M. Heinrich, H. Moya-Cessa, S. Nolte, D.N. Christodoulides and A. Szameit. {\it Phys. Rev. Lett.} {\bf 107}, 103601 (2011).
\bibitem{Chapman2016} R.J. Chapman, M. Santandrea, Z. Huang, G. Corrielli, A. Crespi, M.-H. Yung, R. Osellame and A. Peruzzo. {\it Nat. Comm.} {\bf 7}, 11339 (2016).
\bibitem{Weimann2016} S. Weimann, A. Perez-Leija, M. Lebugle, R. Keil, M. Tichy, M. Grafe, R. Heilmann, S. Nolte, H. Moya-Cessa, G. Weihs, D.N. Christodoulides and A. Szameit. {\it Nat. Comm.} {\bf7}, 11027 (2016).
\bibitem{Blanco2018} A. Blanco-Redondo, B. Bell, D. Oren, B.J. Eggleton and M. Segev. {\it Science} {\bf 362}(6414), 568-571 (2018).
\bibitem{Efremidis2005} N.K. Efremidis and D.N. Christoulides. {Opt. Comm.} {\bf 246}, 345 - 356 (2005).



\bibitem{Meng2004} Y.-C. Meng, Q.-Z. Guo, W.-H. Tan and Z.-M. Huang. {\it JOSA A} {\bf 21}(8), 1518 (2004). 
\bibitem{Perez2013} A. Perez-Leija et al. {\it Phys. Rev. A}, {\bf 87}, 022303 (2013).
\bibitem{Leija2010} A. Perez-Leija, H. Moya-Cessa, A. Szameit and D.N. Christodoulides. {Opt. Lett.}, {\bf 35}(14), 2409-2411 (2010).
\bibitem{Rodriguez2011} B.M. Rodriguez-Lara. {\it Phys. Rev. A}, {\bf 84}, 053845 (2011).
\bibitem{Keil2012} R. Keil, A. Perez-Leija, P. Aleahmad, H. Moya-Cessa, S. Nolte, D.N. Christodoulides and A. Szameit. {\it Opt. Lett.} {\bf 37}, 3801-3803 (2012).



\bibitem{Mollow1967} B.R. Mollow and R.J. Glauber. {\it Phys. Rev.} {\bf 160} (5), 1076-1096 (1967).
\bibitem{Kruse2015} R. Kruse, L. Sansoni, S. Brauner, R. Ricken, C.S. Hamilton, I. Jex and Ch. Silberhorn. {\it Phys. Rev. A} {\bf 92}, 053841 (2015).
\bibitem{Alibart2016} O. Alibart, V. D 'Auria, M. De Micheli, F. Doutre, F. Kaiser, L. Labont{\'e}, T. Lunghi, E. Picholle and S. Tanzilli. {\it J.Opt.} {\bf 18}, 104001 (2016).


\bibitem{Dauria2009} V. D'Auria, S. Fornaro, A. Porzio, S. Solimeno, S. Olivares and M.G.A. Paris. {\it Phys. Rev. Lett.} {\bf 102}, 020502 (2009).
\bibitem{Barral2018} D. Barral, K. Bencheikh, V. D'Auria, S. Tanzilli, N. Belabas and J.A. Levenson. {\it Phys. Rev. A} {\bf 98}, 023857 (2018).
\bibitem{Note0} This can be seen by comparing Figure \ref{F3}a with Figure 2 of ref. \cite{Barral2020}.


\bibitem{Raussendorf2001} R. Raussendorf and H.J. Briegel. {\it Phys. Rev. Lett.} {\bf 86}, 5188 (2001).
\bibitem{Menicucci2006} N.C. Menicucci, P. van Loock, M. Gu, C. Weedbroock, T.C. Ralph and M.A. Nielsen. {\it Phys. Rev. Lett.} {\bf 97}, 110501 (2006).
\bibitem{Gu2009} M. Gu, C. Weedbrook, N.C. Menicucci, T.C. Ralph and P. van Loock. {\it Phys. Rev. A} {\bf 79}, 062318 (2009). 
\bibitem{Ferrini2013} G. Ferrini, J.P. Gazeau, T. Coudreau, C. Fabre and N. Treps. {\it New. J. Phys.} {\bf 15}, 093015 (2013).
\bibitem{vanLoock2003} P. van Loock and A. Furusawa. {\it Phys. Rev. A} {\bf 67}, 052315 (2003).

\bibitem{Ukai2010} R. Ukai, J.-I. Yoshikawa, N. Iwata, P. van Loock and A. Furusawa. {\it Phys. Rev. A} {\bf 81}, 032315 (2010).
\bibitem{Medeiros2014} R. Medeiros de Araujo, J. Roslund, Y. Cai, G. Ferrini, C. Fabre and N. Treps. {\it Phys. Rev. A} {\bf 89}, 053828 (2014).
\bibitem{Beyer2002} H.-G. Beyer and H.-P. Schwefel. {\it Natural Computing} {\bf 1}, 3-52 (2002).
\bibitem{Mondain2019} F. Mondain, T. Lunghi, A. Zavatta, E. Gouzien, F. Doutre, M. De Micheli, S. Tanzilli and V. D'Auria. {\it Photon. Res.} {\bf 7}, A36 (2019).
\bibitem{Takanashi2020} N. Takanashi, T. Kashiwazaki, T. Kazama, K. Enbutsu, R. Kasahara, T. Umeki and A. Furusawa. {\it IEEE J. Quant. Electr.} {\bf 56} (3),6000100 (2020).
\bibitem{Kashiwazaki2020} T. Kashiwazaki, N. Takanashi, T. Yamashima, T. Kazama, K. Enbutsu, R. Kasahara, T. Umeki and A. Furusawa. {\it APL Photon.} {\bf 5}, 036104 (2020).

\bibitem{Boes2018} A. Boes, B. Corcoran, L. Chang, J. Bowers and A. Mitchell. {\it Laser \& Photon. Rev.} {\bf 12}, 1700256 (2018).
\bibitem{Loncar2018} Ch. Wang, C. Langrock, A. Marandi, M. Jankowski, M. Zhang, B. Desiatov, M.M. Fejer and M. Loncar. {\it Optica} {\bf 5} (11), 1438 - 1441 (2018).




\end{thebibliography}
\end{document}